\documentclass[lettersize,journal]{IEEEtran}
\usepackage{amsmath,amsfonts,amssymb}
\usepackage{amsthm}
\usepackage{algorithm}
\usepackage{algpseudocode}
\algnewcommand\Input{\item[\textbf{Input:}]}%
\algnewcommand\Output{\item[\textbf{Output:}]}%
\usepackage{array}
\usepackage{subcaption}
\usepackage{textcomp}
\usepackage{stfloats}
\usepackage{url}
\usepackage{verbatim}
\usepackage{graphicx}
\usepackage{cite}
\usepackage{tabularx}
\usepackage{booktabs}
\usepackage{graphicx}
\usepackage{textcomp}
\usepackage{xcolor}
\usepackage{framed}
\usepackage{diagbox}
\usepackage{url}
\usepackage{multirow}
\usepackage{bm}
\usepackage[cmintegrals]{newtxmath}

\DeclareMathOperator*{\argmin}{arg\,min} 
\newtheorem{theorem}{\bf Theorem}
\newtheorem{corollary}{Corollary}
\newtheorem{lemma}{Lemma}
\newtheorem{proposition}{Proposition}
\newtheorem{obs}{Observation}
\newtheorem{definition}{\bf Definition}

\newtheorem{assumption}{Assumption}

\newtheorem{counter-intuition}{Counter-intuition}

\def\Nsetp{\mathbb{N}_+} 

\def\Ex{\mathbb{E}}

\def\TypeNumber{N}
\def\TypeSet{\mathcal{N}}
\def\TypeIndex{n}
\def\GOperationalcost{c}
\def\GSensingcap{r}
\def\GArrivalprob{p}
\def\OperationalcostN{\GOperationalcost_n}
\def\SensingcapN{\GSensingcap_n}
\def\ArrivalprobN{\GArrivalprob_n}
\def\truncateNum{M}

\def\ActionSet{\mathcal{A}}
\def\BoundActionSet{\hat{\mathcal{A}}}
\def\ReducedActionSet{\mathcal{A}^{RED}}
\def\Action{{A(t)}}
\def\ActionS{{a}}
\def\ActionIndex{i}
\def\OActionNumber{K}
\def\OActionIndex{k}
\def\Actionseq{\mathscr{D}}
\def\OptActionseq{\mathscr{B}}
\def\OptAction{b}

\def\GSuccessprobN{Q}
\def\SuccessprobN{\GSuccessprobN_{\ActionS}}
\def\successprob{success probability}
\def\successprobs{success probabilities}

\def\Successprobs{Success probabilities}

\def\GExpectedrecruitcostN{E}
\def\ExpectedrecruitcostN{\GExpectedrecruitcostN_{\ActionS}}
\def\expectedrecruitcost{expected recruitment cost}

\def\Expectedrecruitcosts{Expected recruitment costs}

\def\MarCosteffectivenessN{\gamma}

\def\Weightedfactor{\beta}
\def\aoi{\delta}

\def\threshold{\theta}
\def\Threshold{\Theta}
\def\state{s}

\def\policy{\phi}
\def\MDP{\Lambda}

\def\cp{company}
\def\Cp{Company}

\def\operationalcost{operational cost}

\def\sensingcap{sensing capability}
\def\sensingcaps{sensing capabilities}

\def\arrivalprob{arrival probability}
\def\arrivalprobs{arrival probabilities}

\def\Tl{Low-type}
\def\Th{High-type}
\def\tl{Low-type}
\def\th{High-type}
\def\Lcost{c_L}
\def\Hcost{c_H}

\def\Lqual{r_L}
\def\Hqual{r_H}
\def\None{$O$}
\def\Ltype{$L$}
\def\Htype{$H$}
\def\Both{$B$}

\def\iterindex{v}
\def\ValueF{V}
\def\PayoffF{J}
\def\IterNum{I}
\def\payoff{{payoff}}
\def\Payoff{{Payoff}}
\def\utility{{gain}}
\def\GGainN{G}
\def\unit{\epsilon}

\ifodd 0
\newcommand{\rev}[1]{{\color{blue}#1}} 
\newcommand{\com}[1]{\textbf{\color{red} (COMMENT: #1) }} 
\newcommand{\comg}[1]{\textbf{\color{green} (COMMENT: #1)}}
\newcommand{\response}[1]{\textbf{\color{green} (RESPONSE: #1)}} 
\else
\newcommand{\rev}[1]{#1}

\newcommand{\com}[1]{}
\newcommand{\comg}[1]{}
\newcommand{\response}[1]{}
\fi

\begin{document}

\title{Efficient and Cost-effective Vehicle Recruitment for HD Map Crowdsourcing}

\author{\IEEEauthorblockN{Wentao Ye, Yuan Luo, Bo Liu, Jianwei Huang
\thanks{Wentao Ye is with the Shenzhen Institute of Artificial Intelligence and Robotics for Society, School of Science and Engineering, The Chinese University of Hong Kong, Shenzhen, Guangdong, 518172, P.R. China. (e-mail: wentaoye@link.cuhk.edu.cn). }
\thanks{Bo Liu is with Shenzhen Institute of Artificial Intelligence and Robotics for Society, Guangdong,  518172, P.R. China (e-mail: liubo@cuhk.edu.cn).}
\thanks{Yuan Luo is with School of Science and Engineering, Shenzhen Key Laboratory of Crowd Intelligence Empowered Low-Carbon Energy Network, The Chinese University of Hong Kong, Shenzhen, Guangdong, 518172, P.R. China. (e-mail: luoyuan@cuhk.edu.cn).}
\thanks{Jianwei Huang is with the School of Science and Engineering, Shenzhen Institute of Artificial Intelligence and Robotics for Society, Shenzhen Key Laboratory of Crowd Intelligence Empowered Low-Carbon Energy Network, and CSIJRI Joint Research Centre on Smart Energy Storage, The Chinese University of Hong Kong, Shenzhen, Guangdong, 518172, P.R. China (email: jianweihuang@cuhk.edu.cn).}
\thanks{This work is supported by the National Natural Science Foundation of China (Project 62472367, 62102343, 62271434, 62203309), Guangdong Basic and Applied Basic Research Foundation (Grant No. 2024A1515011333), Shenzhen Science and Technology Program (Project JCYJ20230807114300001, JCYJ20220818103006012, RCBS20221008093312031),  Shenzhen Key Lab of Crowd Intelligence Empowered Low-Carbon Energy Network (No. ZDSYS20220606100601002), the Shenzhen Stability Science Program 2023, the Shenzhen Institute of Artificial Intelligence and Robotics for Society, Longgang District Shenzhen's "Ten Action Plan" for Supporting Innovation Projects (Grant No. LGKCSDPT2024002, LGKCSDPT2024003) and Guangdong Research (Project 2021QN02X778). (\emph{Corresponding authors: Jianwei Huang and Bo Liu})}
\thanks{Part of the results were published in IEEE WiOpt 2023 \cite{ye2023recruiting}.}
}}

\markboth{Journal of Mobile Computing}%
{Shell \MakeLowercase{\textit{et al.}}: A Sample Article Using IEEEtran.cls for IEEE Journals}

\maketitle

\begin{abstract}
  The high-definition (HD) map is a cornerstone of autonomous driving. The crowdsourcing paradigm is a cost-effective way to keep an HD map up-to-date. Current HD map crowdsourcing mechanisms aim to enhance HD map freshness within recruitment budgets. However, many overlook unique and critical traits of crowdsourcing vehicles, such as random arrival and heterogeneity, leading to either compromised map freshness or excessive recruitment costs. Furthermore, these characteristics complicate the characterization of the feasible space of the optimal recruitment policy, necessitating a method to compute it efficiently in dynamic transportation scenarios.
  To overcome these challenges, we propose an efficient and cost-effective vehicle recruitment (ENTER) mechanism. Specifically, the ENTER mechanism has a threshold structure and balances freshness with recruitment costs while accounting for the vehicles' random arrival and heterogeneity. It also integrates the bound-based relative value iteration (RVI) algorithm, which utilizes the threshold-type structure and upper bounds of thresholds to reduce the feasible space and expedite convergence. Numerical results show that the proposed ENTER mechanism increases the HD map company's \payoff~by \rev{23.40$\%$} and \rev{43.91$\%$} compared to state-of-the-art mechanisms that do not account for vehicle heterogeneity and random arrivals, respectively. Furthermore, the bound-based RVI algorithm in the ENTER mechanism reduces computation time by an average of \rev{$18.91\%$} compared to the leading RVI-based algorithm.
\end{abstract}

\begin{IEEEkeywords}
HD map, Crowdsourcing, Markov decision process, Age of information.
\end{IEEEkeywords}

\section{Introduction}

The high-definition (HD) map is crucial for autonomous driving, providing vehicles with essential static and dynamic environmental information \cite{seif2016autonomous,bao2022high}. In an HD map, dynamic information, such as congestion resolution, incidents (e.g., agglomerate fog), and construction areas, must be highly fresh due to its real-time and unpredictable nature \cite{massow2016deriving}. These updates need to happen within seconds or minutes\cite{b2}. Traditional HD map updates use dedicated mapping vehicles with high-precision sensors like LiDARs and high-accuracy cameras \cite{Ria2015build}. However, maintaining such fleets is costly, restricting their scale and update frequency.

Vehicle-based crowdsourcing has been considered a promising method to update HD maps by both academia \cite{cao2020trajectory,chen2022love,shi2023federated} and industry \cite{Mobileye2017Crowd, Here2021Mer}. For instance, HERE released "HERE HD Live Map" in 2020, using crowdsourcing sensor data to update HD maps from connected vehicles in near-real time. In 2021, HERE HD Live Map became part of Mercedes-Benz's DRIVE PILOT, the first commercially available SAE Level 3 automated driving system \cite{HDLiveMap}. To ensure driver safety and transportation efficiency, HD maps need to be as fresh as possible \cite{hao2022freshness}. Achieving this requires HD map companies (``company" hereafter) to recruit numerous crowdsourcing vehicles (``vehicles" hereafter) and cover their operational costs, which comprise the company’s recruitment costs. While crowdsourcing is often seen as a cheap data source, maintaining an up-to-date HD map can be prohibitively expensive, especially for high-frequency updates and wide coverage \cite{cao2019online}. Thus, designing a mechanism that maximizes the company’s \payoff—defined as the difference between freshness \utility~and recruitment costs—is crucial for the practical success of crowdsourcing-based HD maps\cite{shi2023federated,cheng2023freshness}. However, there are still some issues that need to be addressed.

A key challenge in designing a cost-effective crowdsourcing system is vehicle heterogeneity. Table~\ref{tab:table1} shows that vehicles from different manufacturers have varying sensor configurations, affecting their sensing capabilities and operational costs. Specifically, a vehicle’s sensor suite determines its sensing capability. Processing and transmitting data from different sensors incur varying operational costs due to data sizes, traffic charges, etc \cite{lu2014connected}. Additionally, drivers have different privacy sensitivities \cite{xiong2019privacy}, which influence their willingness to share data and impact operational costs as well. Prior research \cite{huang2019crowdsourcing} indicates that worker heterogeneity in crowdsourcing significantly affects the cost-effectiveness of the crowdsourcing mechanism. Therefore, it is crucial to account for vehicle heterogeneity in our mechanism design.

\begin{table*}[!t]
\caption{Onboard sensor configuration in different vehicle types}
\label{tab:table1}
\centering
\begin{tabular}{ccccc}
\hline
Vehicle type & Camera & LiDAR & MMW radars & Ultrasonic radars\\
\hline
Geely Borui GE  & 5 & $/$ & 1 & 12\\
NIO ES8  & 5 & $/$ & 5 & 12\\
Mercedes-Benz L2 &  5 & $/$ & 4 & 12\\
Xpeng G3 &  8 & $/$ & 3 & 12\\
Tesla &  8 & $/$ & 3 & 12\\
Audi A8 &  5 & 1 & 5 & 12\\
AITO M7 &  11 & 1 & 3 & 12\\
Google Waymo (5th gen) & 29 & 5 & 6 & $/$\\
\hline
\end{tabular}
\end{table*}

The random arrival of vehicles presents a significant challenge. Unlike static sensor-based crowdsourcing systems, vehicle-based crowdsourcing must account for the unpredictable nature of vehicle arrivals. Failure to address this factor can lead to reduced data freshness, ultimately resulting in suboptimal system performance \cite{boubiche2019mobile}. Additionally, the arrival patterns of heterogeneous vehicles can vary \cite{an2021lane}. Considering the random arrival of vehicles adds an extra dimension to vehicle characterization, which also increases the dimensionality of the problem. Therefore, it is both crucial and challenging to consider the diverse and random arrival patterns of vehicles when designing a crowdsourcing system for HD mapping.

Several existing studies optimize the company's \payoff~by balancing freshness \utility~and recruitment costs for HD map updates, but they often overlook some critical issues. Shi \emph{et al.} \cite{shi2023federated} are the first to introduce an overlapping coalition formation game to model vehicle collaboration. However, their work does not consider the important factors of vehicle heterogeneity in operational costs and sensing capabilities.
Cheng \emph{et al.} \cite{cheng2023freshness} do consider vehicle heterogeneity but assume constant vehicle availability, ignoring their random mobility patterns. This neglect leads to sub-optimal freshness and excessive recruitment costs. Our work addresses both freshness \utility~and recruitment costs while accounting for vehicle heterogeneity and random arrivals.

In this paper, we employ the age of information (AoI) to assess the HD map freshness, a metric widely referenced in the literature \cite{kaul2011minimizing,hsu2019scheduling,zhang2021aoi,xu2023aoi,gao2022dynamic}. 
Then, we define the company's freshness \utility~in terms of AoI loss \cite{wang2022dynamic}, which quantifies the profit loss using the outdated HD maps.
In response to AoI dynamics, we apply the Markov decision process (MDP) with infinite states to frame the problem of maximizing the company's \payoff~as widely used in the literature\cite{hsu2019scheduling,bai2023aoi,sombabu2020whittle}. 
However, solving an MDP with infinite states presents a significant challenge. Existing methods \cite{hsu2019scheduling} use a sequence of approximate truncated MDPs to approximate the original MDP. However, this approach only aligns the optimal policies of the truncated and original MDPs as the truncation number approaches infinity. It lacks practical guidance on choosing an appropriate truncation number. In this paper, we introduce a truncated MDP method that maintains the same optimal policy as the original MDP, providing a more practical and definitive solution.

Aside from maximizing the company's \payoff~considering the vehicle's heterogeneity and random arrival, efficiently deriving the optimal recruitment policy for crowdsourcing HD maps is a significant technical challenge. Short-term traffic flow prediction is typically more accurate than long-term prediction \cite{yu2016data,wang2019traffic,majumdar2021congestion}, with "short-term" often referring to just a few minutes \cite{rusyaidi2020review}. Therefore, companies need to derive and apply the optimal policy quickly. However, existing works often overlook this requirement. The relative value iteration (RVI) algorithm, commonly used to solve MDPs \cite{white1963dynamic}, involves solving the Bellman equations, a crucial and time-consuming step \cite{hsu2019scheduling}. With the heterogeneity and random arrival of vehicles, the feasible set of Bellman equations expands exponentially, increasing computation times. To address this challenge, we propose the bound-based RVI algorithm, a time-efficient solution for computing the optimal recruitment policy by leveraging its properties.

Overall, this paper advances the state of the art in the following ways: 
\begin{itemize}
    \item \textit{\texttt{ENTER} Mechanism:} 
    We introduce the \underline{E}fficie\underline{N}t and cos\underline{T}-effective v\underline{E}hicle \underline{R}ecruitment (\texttt{ENTER}) mechanism, the first to efficiently maximize the company's \payoff~by balancing freshness \utility~and recruitment costs for HD maps with heterogeneous, randomly arriving vehicles. Specifically, the \texttt{ENTER} mechanism comprises the optimal vehicle recruitment policy and an efficient algorithm for computing this policy.
    \item \textit{Optimal Policy Analysis:} 
    We prove that the optimal policy is threshold-type age-dependent using MDP modeling. To tackle the challenge of solving an MDP with infinite states, we introduce a criterion for determining the truncation number. This allows us to define a corresponding truncated MDP with finite states, which shares the same thresholds as the original MDP.
    \item \textit{An Efficient Algorithm:} To tackle the challenge of a feasible set in the Bellman equation exponentially growing with the number of vehicle types, we introduce the bound-based RVI algorithm. 
    This algorithm efficiently narrows the feasible set by utilizing the threshold-type structure and computed upper bounds. Consequently, the bound-based RVI algorithm efficiently computes the optimal policy and significantly speeds up computation compared to state-of-the-art RVI-based algorithms.
    \item \textit{Numerical Results:} Numerical results demonstrate that the \texttt{ENTER} mechanism achieves a significant improvement in the company's \payoff. On average, the \texttt{ENTER} mechanism reduces the company's \payoff~by \rev{23.40}$\%$ and $43.91\%$ compared with the state-of-the-art mechanisms \cite{wang2022dynamic} and \cite{cheng2023freshness}, respectively. Moreover, the bound-based RVI algorithm in \texttt{ENTER} reduces the computation time by \rev{$18.91\%$} on average compared to the leading RVI-based algorithm \cite{hsu2019scheduling}. 
\end{itemize}

\section{Related Works}
This section covers two related areas: mechanism design for AoI-related mobile crowdsensing and mechanism design for crowdsourcing-based HD map updates.

\subsection{Mechanism Design for AoI-Related Mobile Crowdsensing}

This section explores mechanisms in mobile crowdsensing that consider the company's \payoff, defined by the difference between freshness \utility~and recruitment costs, for both homogeneous and heterogeneous vehicles.

Wang \emph{et al.} \cite{wang2022dynamic} are the first to propose dynamic pricing for motivating worker contributions and controlling AoI. Xu \emph{et al.} \cite{xu2023aoi} introduced an AoI-guaranteed incentive mechanism (AIM) to effectively address freshness concerns in an incomplete information scenario, taking into account the workers’ social benefits. Gao \emph{et al.} \cite{gao2022dynamic} designed a task pricing scheme to maximize the requester's \payoff, which is defined by the AoI loss and recruitment costs, under the competition of task requesters. However, these approaches assume homogeneous worker capabilities, making them unsuitable for crowdsourcing-based HD maps with heterogeneous vehicles. This limitation prevents the use of unqualified sensing data and reduces frequent update failures, ensuring reliable HD maps and minimizing AoI.

Several works specifically address vehicle heterogeneity in AoI-related mobile crowdsourcing, acknowledging the diverse sensing capabilities and operational costs of vehicles. For example, Wu \emph{i.e.,} \cite{wu2022real} proposed a fair price mechanism that takes into account the non-cooperative relationship among workers, aiming to enhance data freshness.
Cheng \emph{et al.} \cite{cheng2023freshness} designed auction mechanisms to optimize both AoI and data quality within budget constraints. Despite these advances, they overlooked workers' mobility patterns and assumed continuous worker availability, which is inconsistent with the random arrival nature of crowdsourcing vehicles for HD maps. This oversight can lead to large AoI due to the recruitment of unavailable workers.

Our work addresses these gaps by incorporating vehicle heterogeneity and random arrivals while maximizing the company’s \payoff.

\subsection{Mechanism Design for Crowdsourcing-Based HD Map Updates}

This section discusses various mechanisms for updating HD maps through crowdsourcing, focusing on maximizing spatial/temporal coverage under the recruitment budget and minimizing AoI loss under the resource budget. 

Most papers in HD map crowdsourcing optimize spatial/temporal coverage while considering the vehicle's mobility. For instance, Li \emph{et al.} \cite{li2021brief} introduced a periodic crowdsourcing task distribution framework aimed at high time coverage and efficiency with lower recruitment costs. However, this framework assumes uniform rewards for all recruited vehicles, disregarding individual rationality \cite{luo2021budget}. Cao \emph{et al.} \cite{cao2020trajectory} proposed a performance
transfer-based online worker selection scheme to maximize spatial coverage under a recruitment budget, inspired by real-world trajectory analysis. However, they assume that vehicles have homogeneous sensing capabilities. These assumptions overlook vehicle heterogeneity in operational costs and sensing capabilities, potentially leading to unqualified sensing data and the rejection of higher-cost vehicles. Though maximizing both the temporal and spatial coverage is equivalent to minimizing the AoI in some cases \cite{ye2024dual}, these assumptions result in unreliable and outdated HD maps.

While a few papers have optimized freshness utility, vehicle heterogeneity has also been overlooked. For instance, Chen \emph{et al.} \cite{chen2022love} proposed to minimize the AoI loss while considering the computation/communication resource budget and taking into account both the generation and transmission processes of HD maps. Shi \emph{et al.} \cite{shi2023federated} formulated an overlapping coalition formation game to improve the HD map freshness under resource budget, considering the competition among vehicles. It is worth noting that in \cite{chen2022love,shi2023federated}, the term "resource budget" can be interpreted as the recruitment budget when formulating the optimization problem.
Ye \emph{et al.} \cite{ye2024dual} proposed a dual-role AoI-based incentive mechanism (DRAIM) to achieve the tradeoff between freshness \utility~and recruitment costs considering the dual role of vehicles. 
However, these approaches assumed perfect sensing capabilities and did not consider the realistic heterogeneity of vehicle sensing capability. Overcoming this limitation is crucial to avoid unreliable HD maps resulting from the utilization of unqualified sensing data or the occurrence of frequent update failures leading to large AoI. 

Our work addresses these gaps by being the first to maximize the company’s \payoff, defined by freshness \utility~and recruitment costs while considering vehicle heterogeneity in crowdsourcing-based HD map updates. Table~\ref{table:literature} summarizes the key papers in the literature.

\begin{table}[h]
\begin{center}
\caption{\rev{Summary of the Related Work}}
\begin{tabular}{>{\centering\arraybackslash}m{2.0cm} 
				>{\centering\arraybackslash}m{1.0cm} >{\centering\arraybackslash}m{1.0cm} >{\centering\arraybackslash}m{1.8cm} >{\centering\arraybackslash}m{1.0cm}  }
\toprule  
~& Freshness & Cost & Vehicle Heterogeneity & Random Arrival\\[0ex] 
\midrule  
\cite{gao2022dynamic,xu2023aoi,chen2022love}               & $\checkmark$ &$\checkmark$ & $\times$ & $\times$ \\[0ex] 
\cite{wu2022real,cheng2023freshness}                   & $\checkmark$ &$\checkmark$ & $\checkmark$ & $\times$ \\[0ex] 
\cite{li2021brief,cao2020trajectory}  & $\times$     &$\checkmark$ & $\times$ & $\checkmark$ \\[0ex] 
\cite{shi2023federated,wang2022dynamic,ye2024dual}   & $\checkmark$     &$\checkmark$     & $\times$ & $\checkmark$ \\[0ex]
Ours & $\checkmark$    &$\checkmark$ & $\checkmark$ & $\checkmark$  \\[0ex] 
\bottomrule
\end{tabular}
\label{table:literature}
\end{center}
\end{table}

\section{System Model}
In this section, we present the system model aimed at maximizing the company’s \payoff. First, we describe the crowdsourcing paradigm in Section~\ref{subsec: Overview of crowdsourcing in the HD map}, then introduce the {\cp}'s \payoff~function in Section~\ref{subsec: average cost}. Finally, we formulate the company's \payoff~maximization problem as an MDP in Section~\ref{subsec: formulation}.

\subsection{Overview of HD Map Crowdsourcing} \label{subsec: Overview of crowdsourcing in the HD map}
This subsection introduces the crowdsourcing workflow for HD maps.

We consider a company that collects dynamic information at a single PoI\footnote{A more general case is to consider multiple PoIs. However, in many practical scenarios, PoIs can be approximated as independent of each other \cite{bakker2010traffic,sutharsan2020vision,vatchova2023design} due to well-defined traffic patterns. Moreover, our method can be extended to coupled PoIs by incorporating mobility patterns, providing similar insights.}, such as a road junction, from vehicles over a period $t\in\Nsetp$. \rev{The set $\TypeSet=\{\TypeIndex,1\leq \TypeIndex \leq \TypeNumber\}$ represents different vehicle types with heterogeneous distribution of operational costs and sensing capabilities.} \rev{For a type $\TypeIndex$ vehicle, $\mathcal{C}_n$ with mean $\OperationalcostN$ denotes its operational cost distribution, and $\mathcal{R}_n$ with mean $\SensingcapN$ denotes its sensing capability distribution. The sensing capability is the probability of acquiring qualified sensing data primarily influenced by sensor performance.} Note that vehicle's {\sensingcap} refers to the probability of the vehicle acquiring qualified sensing data, which is mainly affected by the sensor performance\footnote{Unqualified sensing data refers to data the company cannot confidently use to update HD maps. For example, for images acquired by cameras, the company can use methods like NR-IQA \cite{kang2014convolutional} and Nanny ML \cite{humphrey2022machine} to assess image quality before using them.}.

Aligning with the standard assumption of the vehicle arrival process in the discrete-time system found in the literature \cite{cai2009adaptive,zhang2018monte,leeftink2023modelling}, we make the following assumption regarding the vehicles' arrival process\footnote{
Many prior studies (\emph{e.g.,} \cite{mrkos2024online,wu2024traffic,haijema2014traffic,amara2023modelling}) used the Bernoulli process to simulate the transportation system. For instance, \cite{mrkos2024online} employed the Bernoulli process to model the arrival of vehicles for electrical vehicle charging management, which is validated by a real-life EV charging station dataset provided by a local German charging station operator. Given this established usage in related research, employing the Bernoulli process in the single PoI system to model the random arrivals of vehicles is well-supported. The more comprehensive Markov Poisson process will be explored in future research when addressing multiple interrelated PoIs.}. 

\begin{assumption}\label{ass: arrival process}
The arrivals of crowdsourcing vehicles of different types are independent and follow Bernoulli processes. Specifically, in each time slot, each vehicle type $\TypeIndex\in\TypeSet$ arrives at the PoI following a Bernoulli process with a probability $p_n$. 
\end{assumption}

We refer to the triplet of parameters $(\ArrivalprobN,\mathcal{C}_n,\mathcal{R}_n)$ as the vehicle parameters.

\begin{figure}[htbp]
    \centering
    \includegraphics[width=7cm]{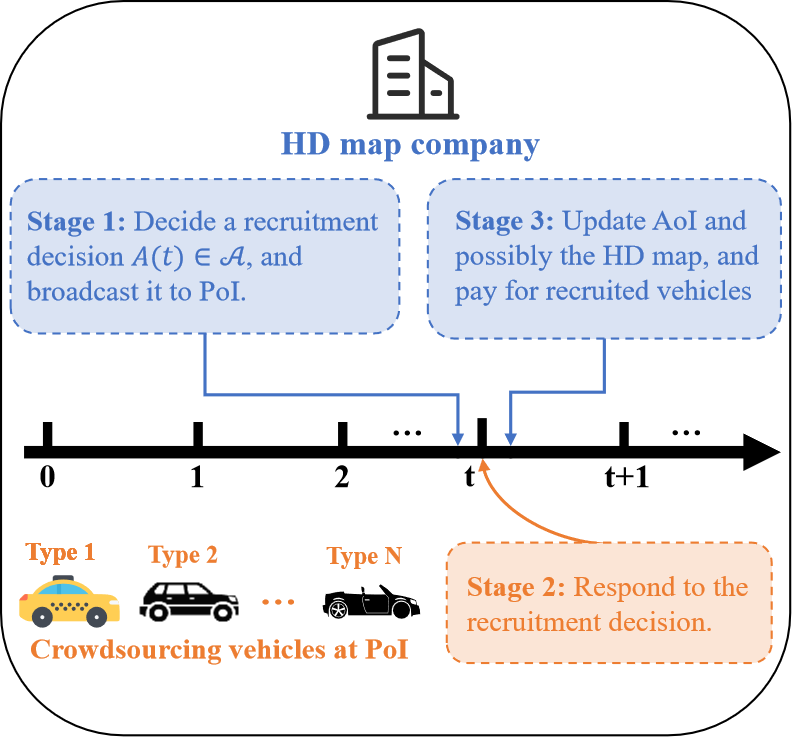}
    \caption{The crowdsourcing workflow for the HD map.}
    \label{fig: crowdsourcing_workflow}
\end{figure}

Fig.~\ref{fig: crowdsourcing_workflow} illustrates the workflow of the HD map crowdsourcing in each time slot $t$. The details are as follows: 
\begin{itemize}
    \item Stage 1: The {\cp} announces a recruitment decision $\Action \in \ActionSet$ at the beginning of the time slot $t$, where $\ActionSet=\mathbb{P}(\TypeSet)$ is the power set of the type set $\TypeSet$ of vehicles\footnote{The power set of a set $A$ is defined as the set of all subsets of the set $A$ including the set itself and the empty set, denoted by $\mathbb{P}(A)$.}. For instance, $\Action=\{\TypeIndex_1, \TypeIndex_2\}$ means recruiting types $\TypeIndex_1$ and $\TypeIndex_2$ vehicles.
    \item Stage 2: Vehicles arrive at the PoI and respond to the company’s recruitment decision for time slot $t$. For instance, if the recruitment decision is $\Action=\{\TypeIndex_1, \TypeIndex_3\}$, only types $\TypeIndex_1$ and $\TypeIndex_3$ vehicles at the PoI will sense the PoI and upload the sensing data to the company. Type $\TypeIndex_2$ vehicles, even if they arrive, will not sense or upload any information.

    \item Stage 3: The {\cp} updates the HD map and the corresponding AoI once receiving the qualified sensing data from recruited vehicles. The company also pays the recruited vehicles once they provide the sensing data\footnote{Because recruited vehicles have already incurred operational costs to acquire and send sensing data. If the {\cp} refuses to reward recruited vehicles that provide unqualified sensing data, she will discourage broad participation.}.
         
\end{itemize}

\subsection{{\Cp}'s \Payoff~Function} \label{subsec: average cost}
In this subsection, we first describe the AoI loss. Next, we introduce the {\cp}'s \payoff~function that depends on the current AoI and her action.

\subsubsection{AoI loss}
AoI increases linearly over time and resets to 1 when the company updates the HD map, as defined in \cite{hsu2019scheduling, pan2021minimizing}. The company updates the HD map when it successfully recruits at least one vehicle that provides qualified sensing data in a time slot. Specifically, the update happens in a time slot $t$ if the {\cp} chooses the recruitment decision $\Action$ and a type $\TypeIndex\in\Action$ vehicle arrives and reports the qualified sensing data.

Considering the AoI at time $t$, denoted as $\aoi(t)$, \rev{the AoI at time $t+1$ will be updated} as follows:

\rev{\begin{equation}\label{Eq: dynamics of AoI}
    \aoi(t+1) = \left\{\begin{array}{ll}
        \aoi(t)+1, & \text{the HD map is not updated in time slot $t$,}\\
        1,  & \text{the HD map is updated in time slot $t$.} \\
    \end{array}
    \right.
\end{equation}}

Similar to the literature \cite{wang2018microeconomic,chen2022love}, AoI loss measures freshness: a higher AoI loss indicates a less fresh HD map. We choose the square function $\aoi^2(t)$, which reflects AoI loss that the {\cp}'s cost convexly increases in the age of its provided information \cite{wang2018microeconomic,chen2022love,chen2023adaptive}.

\subsubsection{\Payoff~function}
\rev{After characterizing AoI loss, we define the company’s freshness \utility~as the negative AoI loss \cite{chen2022love,chen2023adaptive}, \emph{i.e.,} $-\aoi^2(t)$.} The company’s \payoff~is the difference between the expected freshness \utility~and the expected recruitment costs. We focus on the expected values due to the uncertainty introduced by the random arrivals of vehicles and the quality of sensing data.

To address this uncertainty, we define the success probability as the likelihood that at least one vehicle delivers qualified data within a specific time slot. The formal definition is provided below.

\begin{definition}{\textbf{$[$Success Probability $\GSuccessprobN_{\ActionS}$$]$}}
    The success probability $\GSuccessprobN_{\ActionS}$ is the probability that the {\cp} will successfully update the HD map when choosing recruitment decision $\Action=\ActionS\in\ActionSet$.
    \begin{equation} \label{eq: success probability}
        \GSuccessprobN_{\ActionS} = \left\{\begin{array}{ll}
        0,      & \text{if } \ActionS=\emptyset,\\
        1 - \prod_{\TypeIndex\in\ActionS} (1-\SensingcapN\ArrivalprobN),     &  \text{if }  \ActionS\not=\emptyset.\\
        \end{array} 
        \right.
    \end{equation}
\end{definition}

Next, we introduce the expected freshness \utility~and the expected recruitment costs given a recruitment decision $\Action=\ActionS\in\mathcal{A}$. Consider that the current AoI is $\aoi(t)$, then we have the following definitions: 
\begin{itemize}
    \item \textit{Expected freshness \utility} is 
    \begin{eqnarray}\label{eq:expected freshness utiltiy}
    \begin{split}
        \GGainN_{\ActionS}(\aoi(t))=&\left(-\SuccessprobN \cdot 1^2 -(1-\SuccessprobN) (\aoi(t)+1)^2\right)\unit ,\\
        =&\SuccessprobN \left(\aoi^2(t)+2\aoi(t)\right)\unit - (1+\aoi(t))^2\unit,
    \end{split}
    \end{eqnarray}
    where $\epsilon$ is the unit conversion parameter\footnote{The AoI is commonly quantified in temporal units such as seconds and milliseconds. In our paper, we employ seconds. The unit conversion parameter, denoted as $\unit=\$1/s^2$, exemplifies the conversion of seconds squared ($s^2$) to dollars ($\$$). }, and $\SuccessprobN$ is the success probability defined in \eqref{eq: success probability}.
    \item \textit{Expected recruitment costs} is $\ExpectedrecruitcostN$, where 
    \begin{equation}\label{eq:expected recruit cost}
        \GExpectedrecruitcostN_{\ActionS} = \sum_{\TypeIndex\in\ActionS} \ArrivalprobN\OperationalcostN.
    \end{equation}
    Note that if $\ActionS=\emptyset$, $\ExpectedrecruitcostN$ is $0$.
\end{itemize}

A series of recruitment decisions $\Action$ form a recruitment policy $\policy$. Based on the definitions in \eqref{eq:expected freshness utiltiy} and \eqref{eq:expected recruit cost}, we characterize the {\cp}'s \payoff~function $J(\policy)$ under a recruitment policy $\policy$ as follows. 

\begin{equation}\label{eq:payoff function}
    \PayoffF(\policy) = \lim_{T\to\infty}\frac{1}{T}\sum_{t=1}^{T} \beta \GGainN_{\Action}(\aoi(t)) - (1-\Weightedfactor) \GExpectedrecruitcostN_{\Action},
\end{equation}
where $\Weightedfactor$ is the weighted factor balancing the freshness \utility~and recruitment costs. 

\subsection{Problem Formulation} \label{subsec: formulation}
In this subsection, we formulate the company's \payoff~maximization problem as an \emph{average cost MDP} $\MDP$, which is widely used in related works (\emph{e.g.,} \cite{hsu2019scheduling,xu2023aoi,bai2023aoi,fountoulakis2023scheduling,sombabu2020whittle}).

\begin{definition}{\textbf{$[$MDP $\MDP$$]$}}
    The infinite-horizon average cost MDP $\MDP$ is defined as follows.
    \begin{itemize}
        \item \textbf{States}: $S(t)=\aoi(t)\in\mathcal{S}$ representing the AoI at time $t$. Let $\mathcal{S}=\Nsetp$ be the state space, which is countably infinite.
        \item \textbf{Actions}: $\Action\in\ActionSet$, where $\ActionSet=\mathbb{P}(\TypeSet)$ is the action space. In the rest of this paper, we will use the terms ``recruitment decision" and ``action" interchangeably.
        \item \textbf{State transition probability}: $P_{ss'}(\ActionS)$ is the state transition probability from state $S(t)=s$ to $S(t+1)=s'$ under the action $\Action=\ActionS$. If the state $s$ is $\aoi(t)$, then the non-zero $P_{ss'}(a)$ is
        \begin{equation} 
        P_{ss'}(\ActionS)=\left\{
        \begin{array}{ll}          
            Q_{\ActionS}, &\text{if } s'=1,\\
            1-Q_{\ActionS},&\text{if } s'=\aoi(t)+1.\\
        \end{array}\right.
        \end{equation}
        \item \textbf{Immediate cost}: 
        $u(\aoi(t), \ActionS)$ is the cost under state $S(t)=s$ and action $\Action$. If $s=\aoi(t)$ and $\Action=\ActionS$, then the immediate cost $u(\aoi(t), \ActionS)$ is 
        \begin{equation}
            u(\aoi(t), \ActionS)= (1-\Weightedfactor) \GExpectedrecruitcostN_{\ActionS} - \beta \GGainN_{\ActionS}(\aoi(t)).
        \end{equation}
        \item \textbf{Policy}: $\policy=\{\psi_t(\cdot),t\in\Nsetp\}$, where $\psi_t(\cdot)$ is a function mapping from $\mathcal{S}$ to $\mathcal{A}$ at time slot $t$.
        \item \textbf{Value function}: 
        $\ValueF(\policy)$ under a policy $\policy$ is defined as follows\footnote{Given the uncertainty in AoI under the same policy, we denote the expected cost as $\mathbb{E}^\policy$, acknowledging that AoI evolves randomly under policy $\policy$.}:
        \begin{equation}\label{eq:objective function}
            \ValueF(\policy)= \lim_{T\to\infty} \frac{1}{T} \mathbb{E}^\policy\left[\sum_{t=1}^T u\left(S(t),\psi_t(S(t))\right)\right].
        \end{equation}
    \end{itemize}
\end{definition}

The objective of MDP $\MDP$ is to minimize the value function $\ValueF(\policy)$ in \eqref{eq:objective function}, which is equivalent to maximizing the company’s \payoff~function $\PayoffF(\policy)$ since $\ValueF(\policy)=-\PayoffF(\policy)$. By computing the optimal policy $\policy^*$, we can achieve the minimum value $\ValueF^*$ and the maximum \payoff~$\PayoffF^*(=-\ValueF^*)$, simultaneously.

Next, we will limit our consideration to deterministic stationary policies based on Lemma~\ref{lemma:deterministic stationary optimal policy}.

\begin{lemma}\label{lemma:deterministic stationary optimal policy}
    There exists an optimal policy of MDP $\MDP$ that is deterministic stationary.
\end{lemma}

The proof of Lemma~\ref{lemma:deterministic stationary optimal policy} can be found in Appendix A. Based on Lemma~\ref{lemma:deterministic stationary optimal policy}, identifying the optimal deterministic stationary policy is sufficient for computing the optimal policy $\policy^*$ of MDP $\MDP$. Therefore, we can treat the optimal policy $\policy^*$ as a function that maps from each state to a single action.

There are two challenges in solving the formulated MDP: the infinite states within the MDP and the efficient derivation of the optimal policy. In the following section, we introduce a truncated MDP and an efficient algorithm that leverages the properties of the optimal policy to address these issues, respectively.

\section{\rev{ENTER Mechanism Design}}\label{sec: main results}
In this section, we introduce the \texttt{ENTER} mechanism, which consists of the optimal recruitment policy and the bound-based RVI algorithm. In Section~\ref{subsec: Characterization of the optimality}, we present the properties of the optimal policy, including the threshold-type structure and the upper bounds of thresholds. 
In Section~\ref{subsec: approximate MDP}, we further address the challenge of solving the MDP with infinite states by introducing a truncated MDP that shares the same thresholds.
Finally, in Section~\ref{subsec: BRVIA}, we propose the bound-based RVI algorithm, which efficiently computes the optimal policy using the threshold-type structure and upper bounds of thresholds.

\subsection{Characterization of the Optimal Policy Property}\label{subsec: Characterization of the optimality}
This subsection characterizes the optimal policy's structure type and then determines the action order in the optimal policy. Finally, we present the upper bounds of thresholds.

\subsubsection{Structure type} We will show that the optimal policy has a threshold-type and age-dependent structure, providing valuable insights for determining the optimal policy. To understand this structure, we first define the \emph{threshold-type and age-dependent policy}.

\begin{definition}{\textbf{$[$Threshold-type Age-dependent Policy$]$}}
    A deterministic stationary policy $\policy$ is a threshold-type age-dependent policy of the MDP $\MDP$ if there exist a finite $I$-element sequence of actions $\left(\ActionS_i,1\leq i \leq I\right)$ and a set of $I-1$ thresholds $\Threshold=\left\{\threshold_{\ActionS_i\to \ActionS_{i+1}}\in\Nsetp | \ActionS_i\not=\ActionS_{i+1}, \ActionS_i\in\ActionSet,\ActionS_{i+1}\in\ActionSet,\forall 1\leq i < I\right\}$ such that 
    \begin{eqnarray}\label{def:deterministic stationary policy}
        \policy(\state) = \left\{\begin{array}{ll}
        \ActionS_1, & \text{if } \state < \threshold_{\ActionS_1\to \ActionS_{2}}, \\
        \ActionS_{i}, & \text{if } \threshold_{\ActionS_{i-1}\to \ActionS_{i}} \leq \state < \threshold_{\ActionS_{i}\to \ActionS_{i+1}}, \forall 1 < i < I, \\
        \ActionS_{I}, & \text{if } \state \geq \threshold_{\ActionS_{I-1}\to \ActionS_{I}}.\\
        \end{array}\right.
    \end{eqnarray}
\end{definition}

The sequence of actions $\left(\ActionS_i,1\leq i \leq I\right)$ in Definition~\ref{def:deterministic stationary policy} is the action order the policy $\policy$ takes. The action order enables us to compute the upper bounds of thresholds, which is critical for designing the bound-based RVI algorithm. Similar to \cite{hsu2019scheduling}, we derive the structure type of the optimal policy as follows. 

\begin{theorem}\label{thm:optimal policy}
    For MDP $\MDP$, there exists an optimal threshold-type age-dependent policy $\policy^*$. Considering                $K\in\Nsetp$ different actions in the optimal policy, the optimal policy $\policy^*$ has a finite action order $\OptActionseq=\left(\OptAction_{\OActionIndex}\in\ActionSet, 1 \leq \OActionIndex \leq \OActionNumber\right)$ and a set of $K-1$ thresholds $$\Threshold=\left\{\threshold_{\OptAction_\OActionIndex \to \OptAction_{\OActionIndex+1}}\in\Nsetp | \OptAction_\OActionIndex\not= \OptAction_{\OActionIndex+1},\forall 1\leq \OActionIndex < \OActionNumber\right\}.$$
\end{theorem}
 
The proof of Theorem~\ref{thm:optimal policy} is shown in Appendix B.
Theorem~\ref{thm:optimal policy} reveals the structure type of the optimal policy, enabling efficient computation using action order and upper bounds of thresholds. Section~\ref{subsubsec: action order} illustrates how to derive the action order of the optimal policy $\OptActionseq$, while Section~\ref{subsec: BRVIA} covers the derivation of thresholds $\Threshold$.

\subsubsection{Action order} \label{subsubsec: action order} 
To determine the action order in the optimal policy, we first derive a key property in Proposition~\ref{prop:decsendingQ}. We then introduce the concept of \emph{marginal cost-effectiveness} to assess the efficiency of action transitions with AoI growth. Finally, we propose an algorithm to determine the optimal policy’s action order based on these principles.

We first show that the action order of the optimal policy follows the order of descending \successprob, as follows.
\begin{proposition}\label{prop:decsendingQ}
    In the action order $\OptActionseq=\left(\OptAction_{\OActionIndex}\in\ActionSet, 1 \leq \OActionIndex \leq \OActionNumber\right)$ of the optimal policy $\policy^*$, for  $1\leq \OActionIndex < \OActionNumber$, actions $\OptAction_\OActionIndex$ and $\OptAction_{\OActionIndex+1}$ follow a descending \successprob, i.e., $\GSuccessprobN_{\OptAction_{\OActionIndex}} \leq \GSuccessprobN_{\OptAction_{\OActionIndex+1}}.$
\end{proposition}

Proposition~\ref{prop:decsendingQ} implies that the company has a higher preference for updating the HD map as the AoI grows. This is because as AoI increases, the AoI loss occupies the dominant position gradually in the company's \payoff. 
The proof of Proposition~\ref{prop:decsendingQ} is in Appendix C. 
Without loss of generality, we assume that the \successprob~of action $\ActionS_i$ increases with its index $i$. 
For clarity, let us introduce a sequence of actions with ascending \successprobs, denoted as $\Actionseq=\left(\ActionS_{\ActionIndex}, 1 \leq \ActionIndex \leq 2^\TypeNumber\right)$, where $\GSuccessprobN_{\ActionS_i}<\GSuccessprobN_{\ActionS_{i+1}}$ for all $1\leq i < 2^\TypeNumber$. 
Based on Proposition~\ref{prop:decsendingQ}, the action order $\OptActionseq$ of the optimal policy $\policy^*$ must be the subsequence of $\Actionseq$.

To determine the optimal policy structure, we define the \emph{marginal cost-effectiveness} to assess the cost-effectiveness of a company's action change as follows.
\begin{definition}{\textbf{$[$Marginal Cost-effectiveness $\MarCosteffectivenessN_{\ActionS_i,\ActionS_j}$$]$}}
    The marginal cost-effectiveness of changing action from $\ActionS_i$ to $\ActionS_j$ is the ratio between the increase of {\expectedrecruitcost} and the increase of {\successprob}: 
    \begin{equation}
        \MarCosteffectivenessN_{\ActionS_i,\ActionS_j} = \frac{\GExpectedrecruitcostN_{\ActionS_j}-\GExpectedrecruitcostN_{\ActionS_i}}{\GSuccessprobN_{\ActionS_j}-\GSuccessprobN_{\ActionS_i}}.
    \end{equation}
\end{definition}

A smaller value of marginal cost-effectiveness indicates that the company is more inclined to alter her action. For instance, if $\GSuccessprobN_{\ActionS_j}>\GSuccessprobN_{\ActionS_i}$ and $\MarCosteffectivenessN_{\ActionS_i,\ActionS_j}$ is small, then the company is likely to opt for action $\ActionS_j$ instead of $\ActionS_i$. If $\ActionS_i=\emptyset$, we simply refer to $\MarCosteffectivenessN_{\ActionS_i,\ActionS_j}$ as the \emph{cost-effectiveness} of $\ActionS_j$ for short. 

Next, we compute the optimal policy's action order by integrating Proposition~\ref{prop:decsendingQ} with the concept of marginal cost-effectiveness, as shown in Proposition~\ref{prop:optimal recruitment order equation}.

\begin{proposition}\label{prop:optimal recruitment order equation}
    In the action order $\OptActionseq=\{\OptAction_{\OActionIndex}, 1 \leq \OActionIndex \leq \OActionNumber\}$ of the optimal policy $\policy^*$, for any $\OActionIndex$, the following equation holds:
    \begin{equation}\label{eq:}
        \OptAction_{\OActionIndex+1} = \argmin_{j\in\left\{\ActionIndex|\GSuccessprobN_{\ActionS_{\ActionIndex}}>\GSuccessprobN_{\OptAction_{\OActionIndex}}\right\}}\frac{\GExpectedrecruitcostN_{\ActionS_{j}}-\GExpectedrecruitcostN_{\OptAction_{\OActionIndex}}}{\GSuccessprobN_{\ActionS_{j}}-\GSuccessprobN_{\OptAction_{\OActionIndex}}}.
    \end{equation}
\end{proposition}

The proof of Proposition~\ref{prop:optimal recruitment order equation} is present in Appendix D. 
Proposition~\ref{prop:optimal recruitment order equation} enables us to propose Algorithm~\ref{alg:structure determine} to compute the action order of the optimal policy. Specifically, as AoI continues to grow, the algorithm chooses the action that yields the minimum increase in marginal cost-effectiveness relative to the current action, as shown in Line 4 of Algorithm~\ref{alg:structure determine}. Understanding the optimal action order derived from Algorithm~\ref{alg:structure determine} allows us to efficiently compute the optimal recruitment policy for HD map crowdsourcing in Section~\ref{subsec: BRVIA}. 

\begin{algorithm}[!t]
\caption{Optimal policy structure determination}
\begin{algorithmic}[1] 
\Input{Sorted action sequence: $\mathcal{A}$
\Statex \quad \, \Successprobs: $\{\SuccessprobN,\ActionS\in\ActionSet\}$
\Statex \quad \, \Expectedrecruitcosts: $\{\ExpectedrecruitcostN,\ActionS\in\ActionSet\}$}
\Output{The optimal action order: $\OptActionseq$}
\State \textsc{Initialize} $\OptActionseq \leftarrow\{\OptAction_1\leftarrow\ActionS_1\}$
\State $i \leftarrow 1$, $j \leftarrow 2$
\While{$\ActionIndex \leq 2^{\TypeNumber}$}
    \State $\ActionS_{j^*} \leftarrow \argmin_{j>\ActionIndex} \MarCosteffectivenessN_{\ActionS_{\ActionIndex},\ActionS_{j}}$
    \State Add $\ActionS_{j^*}$ in $\OptActionseq$, i.e., $\OptAction_{\OActionIndex+1}\leftarrow\ActionS_{j^*}$
    \State $\OActionIndex\leftarrow\OActionIndex + 1$, $\ActionIndex \leftarrow j^*$
\EndWhile
\end{algorithmic}
\label{alg:structure determine}
\end{algorithm}

\subsubsection{Upper bounds of thresholds} Finally, we compute the upper bounds of thresholds corresponding to actions in action order $\OptActionseq$ of the optimal policy in Lemma~\ref{lemma:upper bounds}.

\begin{lemma} \label{lemma:upper bounds}
    Given the optimal action order $\OptActionseq$, for $2\leq \OActionIndex \leq \OActionNumber$, the optimal threshold $\threshold_{\OptAction_{\OActionIndex-1}\to\OptAction_{\OActionIndex}}$ is upper bounded by the following value
    \begin{equation}
        \hat{\threshold}_{\OptAction_{\OActionIndex-1}\to\OptAction_{\OActionIndex}} = \left\lceil\sqrt{\frac{1}{\unit}+\frac{1-\Weightedfactor}{\Weightedfactor\unit}\MarCosteffectivenessN_{\OptAction_{\OActionIndex-1}, \OptAction_{\OActionIndex}}}-1\right\rceil.
    \end{equation}
\end{lemma}

The proof of Lemma~\ref{lemma:upper bounds} is detailed in Appendix E. 
Though we derive the action order of the optimal policy, MDP $\MDP$ is challenging to analyze due to its infinite-state space. In Section~\ref{subsec: approximate MDP}, we will address this challenge.

\subsection{Truncated MDP with Finite States} \label{subsec: approximate MDP}
This subsection introduces a criterion for the truncation number determination and proposes the corresponding truncated MDP to address the challenge of solving MDPs with an infinite-state space.

The classical method for solving an MDP is to apply an RVI or RVI-based algorithm \cite{white1963dynamic,puterman2014markov}. However, updating an infinite number of states in each iteration renders these algorithms impractical for our problem. 
Existing approaches (\emph{e.g.,} \cite{hsu2019scheduling}) propose a sequence of approximate truncated MDPs that converge to the original MDP as the truncation number approaches infinity. Based on this convergence, these methods establish the connection between the optimality of the approximate truncated MDP and the original MDP. However, it is also challenging to determine the proper number of truncated states. 
In this paper, we propose an exact truncated MDP that shares the same optimal policy as the original MDP and determine the value of the truncation number. This exact truncation ensures practical and efficient solutions.

Let M denote the number of truncated states\footnote{Here, we reasonably choose the number of truncated states to be greater than the upper bound of the largest threshold, \emph{i.e.,} $\truncateNum>\hat{\threshold}_{\OptAction_{K-1}\to\TypeSet}$. Later, we provide reasons for the choice in Theorem~\ref{thm:approximate MDP}.}.  We define the truncated AoI using the notation $[x]_{\truncateNum}^+$, as follows:

\begin{eqnarray*}
    [x]_{\truncateNum}^+ = \left\{
        \begin{array}{ll}
            x, & \emph{if } x<{\truncateNum},\\
            \truncateNum, & \emph{if } \text{otherwise}.
        \end{array}
    \right.
\end{eqnarray*}

The dynamics of the truncated AoI $\aoi^{\truncateNum}(t)$, mirror those in \eqref{Eq: dynamics of AoI}, except when the HD map is not updated:
\begin{equation*}
    \aoi^{\truncateNum}(t+1) = \left\{\begin{array}{ll}        
        \left[\aoi(t)+1\right]^+_{\truncateNum}, & \text{if the HD map is not updated},\\
        1,  & \text{if the HD map is updated.} \\
    \end{array}
    \right.
\end{equation*}

Let $\mathcal{S}^{\truncateNum}=\{n\leq \truncateNum|n\in\Nsetp\}$ be the truncated state space, which is countably finite. We define the \emph{truncated MDP} $\MDP^{\truncateNum}$ as follows.

\begin{definition}{\textbf{$[$Truncated MDP $\MDP^{\truncateNum}$$]$}} \label{def:truncated MDP}
    The truncated MDP $\MDP^{\truncateNum}$ is same as the original MDP $\MDP$ except for
    \begin{itemize}
        \item \textbf{States}: $S^{\truncateNum}(t)=\aoi^{\truncateNum}(t)\in\mathcal{S}^{\truncateNum}$, \emph{i.e.,} $\aoi^{\truncateNum}(t)$ is the truncated AoI at that slot.            
        \item \textbf{State transition probability}: $P^{\truncateNum}_{ss'}(\ActionS)$, \emph{i.e.,} the state transition probability from state $S^{\truncateNum}(t)=s$ to $S^{\truncateNum}(t+1)=s'$ under the action $\Action=\ActionS$. Consider the state $s$ is $\aoi^{\truncateNum}(t)$, the non-zero $P^{\truncateNum}_{ss'}(\ActionS)$ is\\
        \begin{equation} 
        P^{\truncateNum}_{ss'}(\ActionS)=\left\{
        \begin{array}{ll}        
            Q_\ActionS, &\text{if } s'=1,\\
            1-Q_\ActionS,&\text{if } s'=[\aoi^{\truncateNum}(t)+1]_{\truncateNum}^+.\\
        \end{array}\right.
        \end{equation}
    \end{itemize}
\end{definition}

Therefore, truncated MDP $\MDP^{\truncateNum}$ has finite states and bounded immediate costs since the AoI is truncated.

Finally, we establish the optimality connection between the truncated MDP $\MDP^{\truncateNum}$ and the original MDP $\MDP$. Let $\IterNum$ denote the iteration number at which the RVI-based algorithm converges.

\begin{theorem} \label{thm:approximate MDP}
     The optimal policy of the truncated MDP $\MDP^{\truncateNum}$ shares the same thresholds as the optimal policy of the original MDP $\MDP$, if 
     \begin{eqnarray}\label{eq:optimality condition}
         \truncateNum \geq \threshold_{\OptAction_{K-1}\to\TypeSet}+\IterNum.
     \end{eqnarray}
\end{theorem}

Theorem~\ref{thm:approximate MDP} provides a criterion for the appropriate truncation number, crucial for the practical success of HD map crowdsourcing. Setting a truncation number that is too large can excessively prolong computation times, failing to meet the dynamic requirements of traffic systems. Conversely, a truncation number that is too small results in a mismatch between the optimal policies of the truncated MDP $\MDP^{\truncateNum}$ and the original MDP $\MDP$, thereby reducing the company's \payoff. 
The proof of Theorem~\ref{thm:approximate MDP} can be found in Appendix F.

We want to further elaborate on Theorem~\ref{thm:approximate MDP}. First, the number of iterations, $\IterNum$, is variable and depends on the parameters of the RVI-based algorithm, such as the convergence criteria. Therefore, it cannot be predetermined. Second, the threshold $\threshold_{\OptAction_{K-1}\to\TypeSet}$ is not known in advance. However, we can start with a reasonable large initial truncation number, $\truncateNum=\hat{\threshold}{\OptAction{K-1}\to\TypeSet}$, and verify if \eqref{eq:optimality condition} is met. If this condition is satisfied, the optimal policy has been identified. If not, we adjust $\truncateNum$ and rerun the algorithm until \eqref{eq:optimality condition} is met.

There are many RVI-based algorithms for computing the optimal policy of truncated MDP $\MDP^{\truncateNum}$, such as the structural RVI (SRVI) [20]. However, they are inefficient due to the time-intensive minimization operation required for all states $S^{\truncateNum}(t)=s$ in each iteration:

\begin{equation}\label{eq: optimal policy update}
    \begin{split}
    &\ValueF_{\iterindex+1}\left(s\right)=\min_{\ActionS\in\mathcal{A}} \left[ u\left(s,\ActionS\right) + \sum_{s'\in\mathcal{S}^{\truncateNum}} P^{\truncateNum}_{ss'}(\ActionS) \ValueF_{\iterindex}\left(s'\right) - \ValueF_{\iterindex}(1)\right].
    \end{split}
\end{equation}

This step is one of the most time-consuming parts of RVI-based algorithms \cite{hsu2019scheduling}. The computational complexity for all truncated states is $O(|\mathcal{A}|\times \truncateNum)$, derived from the size of the feasible set and the number of truncated states. As the size of the feasible set in \eqref{eq: optimal policy update} exponentially increases with the number of vehicle types, the complexity of RVI-based algorithms also grows exponentially.

In the next subsection, we propose the bound-based RVI algorithm to solve truncated MDP $\MDP^{\truncateNum}$ efficiently. 

\subsection{Bound-based Relative Value Iteration Algorithm} \label{subsec: BRVIA}

In this subsection, we develop the bound-based RVI algorithm, which efficiently computes the optimal policy by reducing the feasible set of the Bellman equation. Specifically, we leverage the upper bounds of thresholds and the threshold-type structure to narrow this feasible set, which will significantly reduce computation time.

\subsubsection{Through threshold upper bounds} Based on the upper bounds of thresholds in the optimal policy characterized by Lemma~\ref{lemma:upper bounds}, we condense the feasible set $\mathcal{A}$ in \eqref{eq: optimal policy update} for faster iterations. Specifically, we define the \emph{bounded feasible set} $\BoundActionSet(\state)$ as follows:
\begin{equation}\label{eq: feasible action space function} 
\BoundActionSet(\state) = \left\{ \begin{array}{ll}
        \OptActionseq, & \emph{if } \state < \hat{\threshold}_{\OptAction_{1}\to\OptAction_{2}},\\
        \{\OptAction_j |\OActionIndex-1\leq j \leq \OActionNumber\}, & \emph{if }  \hat{\threshold}_{\OptAction_{\OActionIndex-2}\to\OptAction_{\OActionIndex-1}}\leq \state < \hat{\threshold}_{\OptAction_{\OActionIndex-1}\to\OptAction_{\OActionIndex}},\\
        \TypeSet , &\emph{if }  \state \geq \hat{\threshold}_{\OptAction_{\OActionNumber-1}\to\OptAction_{\OActionNumber}}.\\
    \end{array}
\right.
\end{equation}

In addition to using the derived threshold upper bounds, we can further streamline the feasible set by applying the threshold-type property of policies during iterations.

\subsubsection{Through structural property of the optimal policy}

The RVI-based algorithm iterations compute a policy that maintains the threshold-type structure, as shown in Lemma~\ref{lemma:iterative structural property}. This structure enables us to tighten the feasible set and expedite iterations effectively. Specifically, we can eliminate sub-optimal actions in each iteration by comparing their success probabilities with the optimal action from the previous state.

\begin{lemma}\label{lemma:iterative structural property}
    Consider that $\policy^{(v)}$ is the policy computed by the Bellman equation in interaction $v$. For any states $s_1<s_2\in\mathcal{S}^{\truncateNum}$,
    \begin{eqnarray}
        \GSuccessprobN_{\policy^{(v)}(s_1)} \leq \GSuccessprobN_{\policy^{(v)}(s_2)}.
    \end{eqnarray}
\end{lemma}

The proof of Lemma~\ref{lemma:iterative structural property} is based on the Bellman equation used in the value iterations, as detailed in Appendix G. 
Based on Lemma~\ref{lemma:iterative structural property} and the bounded feasible set, we introduce the \emph{reduced feasible set $\ReducedActionSet$}, as defined in \eqref{eq:Reduced feasible set}, to replace the original feasible set. 
\begin{definition}{\textbf{$[$Reduced Feasible Set $\ReducedActionSet$$]$}}
    Consider that the current state is $S(t)=\state$, and the optimal action of the last state $\state-1$ is $\policy^*(\state-1)$. The reduced feasible set $\ReducedActionSet$ of the current state is
    \begin{eqnarray} \label{eq:Reduced feasible set}
        \mathcal{A}^{RED}=\BoundActionSet(\state) - \{a|\GSuccessprobN_{\ActionS}<\GSuccessprobN_{\policy^*(\state-1)}\}.
    \end{eqnarray}
\end{definition}

\rev{The rationale behind the reduced action space is removing the impossible actions from the feasible set. Specifically, we prove that the optimal policy has a threshold-type structure, where the success probability of the action in the optimal policy increases in the AoI, as indicated in Theorem 1 and Proposition 1. Therefore, we can exclude actions with success probabilities exceeding the upper bound and falling below that of the optimal action in the previous state.}

Employing the reduced feasible set $\ReducedActionSet$, we enhance the RVI-based algorithm efficiency in the next subsubsection. 

\subsubsection{Bound-based RVI algorithm} We propose the bound-based RVI algorithm to reduce the computation time of the RVI-based algorithms by utilizing the reduced feasible set $\ReducedActionSet$. This algorithm starts with the determination of the optimal policy structure (\emph{i.e.,} Line 1), and then computes the bounded feasible set based on the threshold upper bounds (\emph{i.e.,} Line 2). Algorithm~\ref{alg:BRVIA} initializes and backs up the value function before iteration (Lines 3-4). The iteration process runs from Lines 6 to 12, repeating the following procedures until the value function converges (\emph{i.e.,} Line 5). 
\begin{itemize}
    \item Line 6: Back up the value function.
    \item Lines 7-11: Compute the value function. Specifically, update the reduced feasible set $\ReducedActionSet$ (Line 8), solve the Bellman equation where the original feasible set $\ActionSet$ is replaced by the $\ReducedActionSet$ (Line 9), and compute the relative value of each state (Line 10).
    \item Line 12: Update the value function in a batch.
\end{itemize}

\begin{algorithm}
\caption{Bound-based RVI algorithm}
\label{alg:BRVIA}
\begin{algorithmic}[1]
\Input{Weighted factor: $\Weightedfactor$
\Statex \quad \, Vehicle parameters: $(\ArrivalprobN,\OperationalcostN,\SensingcapN),\forall \TypeIndex\in\TypeSet$
\Statex \quad \, Convergence tolerance: $\epsilon$
\Statex \quad \, Number of truncated states: $\truncateNum$ }
\Output{Optimal policy $\policy^*$} 
\State Determine the structure of the optimal policy based on Algorithm~\ref{alg:structure determine}
\State Calculate the bounded feasible set $\BoundActionSet(\state)$ based on \eqref{eq: feasible action space function} for all truncated states $\state\in\mathcal{S}^{\truncateNum}$
\State Initialize $\ValueF^{\truncateNum}(\state) \leftarrow 0$ for all truncated states $\state\in\mathcal{S}^{\truncateNum}$
\State Use $\ValueF^{\truncateNum}_{last}(\state)$ to back up $\ValueF^{\truncateNum}$ of the last round
\While{$\left|\frac{\ValueF^{\truncateNum}-\ValueF^{\truncateNum}_{last}}{\ValueF^{\truncateNum}}\right|>\epsilon$}
    \State $\ValueF^{\truncateNum}_{last}(\state) \leftarrow \ValueF^{\truncateNum}(\state)$ for all $\state\in\mathcal{S}^{\truncateNum}$
    \For{$\state\in\mathcal{S}^{\truncateNum}$}
        \State Update reduced feasible set $\mathcal{A}^{RED}$ based on \eqref{eq:Reduced feasible set}
        \State $\policy^*(\state) \leftarrow \argmin_{a\in\mathcal{A}^{RED}} u(\state,\ActionS)+\Ex[\ValueF^{\truncateNum}(\state')]$
        \State $\ValueF^{\truncateNum}_{tmp}(\state)\leftarrow u(\state,\policy^*(\state))+\mathbb{E}[\ValueF^{\truncateNum}(\state')]-\ValueF^{\truncateNum}(1)$
    \EndFor
    \State $\ValueF^{\truncateNum}(\state)\leftarrow \ValueF^{\truncateNum}_{tmp}(\state)$ for all $\state\in\mathcal{S}^{\truncateNum}$
\EndWhile
\end{algorithmic}
\end{algorithm}

Next, we establish the optimality of the bound-based RVI algorithm for truncated MDP $\MDP^{\truncateNum}$.

\begin{theorem} \label{thm:alg converges}
    The output $\policy^*$ in Algorithm~\ref{alg:BRVIA} converges to the optimal policy of truncated MDP $\MDP^{\truncateNum}$ in a finite number of iterations.
\end{theorem}

The proof of Theorem~\ref{thm:alg converges} can be found in Appendix H. 
Combining Theorems~\ref{thm:approximate MDP} and \ref{thm:alg converges}, we conclude that Algorithm~\ref{alg:BRVIA} can compute the optimal policy of MDP $\MDP$.

In summary, the optimal policy of the MDP outlined in Theorem~\ref{thm:optimal policy} and the proposed bound-based RVI algorithm we proposed form the \texttt{ENTER} mechanism for HD map crowdsourcing.

\section{Numerical Results}\label{Sec: simulations}

In this section, we perform extensive numerical experiments to showcase the superior cost-effectiveness and efficiency of the proposed \texttt{ENTER} mechanism \rev{(including the optimal recruitment policy and bound-based RVI algorithm)} in Sections~\ref{subsec: comparison of the average cost under different policies} and \ref{subsec: efficiency comparison}. We also explore how vehicle parameters influence the optimal policy in the \texttt{ENTER} mechanism, demonstrating its sensitivity to these parameters in Section~\ref{subsec:para sensitivity}. 

\addtolength{\topmargin}{0.16in}

\subsection{Cost-effectiveness Comparison} \label{subsec: comparison of the average cost under different policies}
In this subsection, we compare the \texttt{ENTER} mechanism to three baselines: zero-wait policy, DRAIM, and auction mechanism.

\begin{itemize}
    \item Zero-wait policy: A naive policy that ignores recruitment costs in the company’s \payoff~function, \emph{i.e.,} recruiting vehicles of all types at all times.
    \item DRAIM \cite{ye2024dual}: This mechanism maximizes the same tradeoff as the ENTER mechanism but ignores vehicles’ heterogeneous sensing capabilities.
    \item Auction mechanism \cite{cheng2023freshness}: This policy considers the same tradeoff as the \texttt{ENTER} mechanism but overlooks the random arrivals of vehicles at the PoI.
\end{itemize}

\rev{Except for the company's \payoff, we examine the average AoI and recruitment cost, defined as follows.
\begin{definition}{\textbf{$[$Average AoI $\bar{\aoi}$ and Recruitment Cost $\bar{c}$ $]$}}
    Let the stationary distribution of the MDP $\MDP$ under the optimal policy be represented by $\pi^*(\state)$. The average AoI is calculated as: 
    \begin{eqnarray}
        \bar{\aoi} = \sum_{\state\in\mathcal{S}} \pi^*(s) \cdot s.
    \end{eqnarray}
    The average recruitment cost is determined by:
    \begin{eqnarray}
        \bar{c} = \sum_{\state\in\mathcal{S}} \pi^*(s) \cdot \GExpectedrecruitcostN_{\policy^*(s)}.
    \end{eqnarray}
\end{definition}

}
We conducted two numerical experiments to demonstrate the  \texttt{ENTER} mechanism's superiority, assessing how the weighted factor and the number of vehicle types affect the {\cp}'s \payoff.

\subsubsection{The impact of the weighted factor on the company's \payoff}
\begin{figure}[htbp]
    \centering
    \vspace{-0.4cm}
    \includegraphics[width=.66\linewidth]{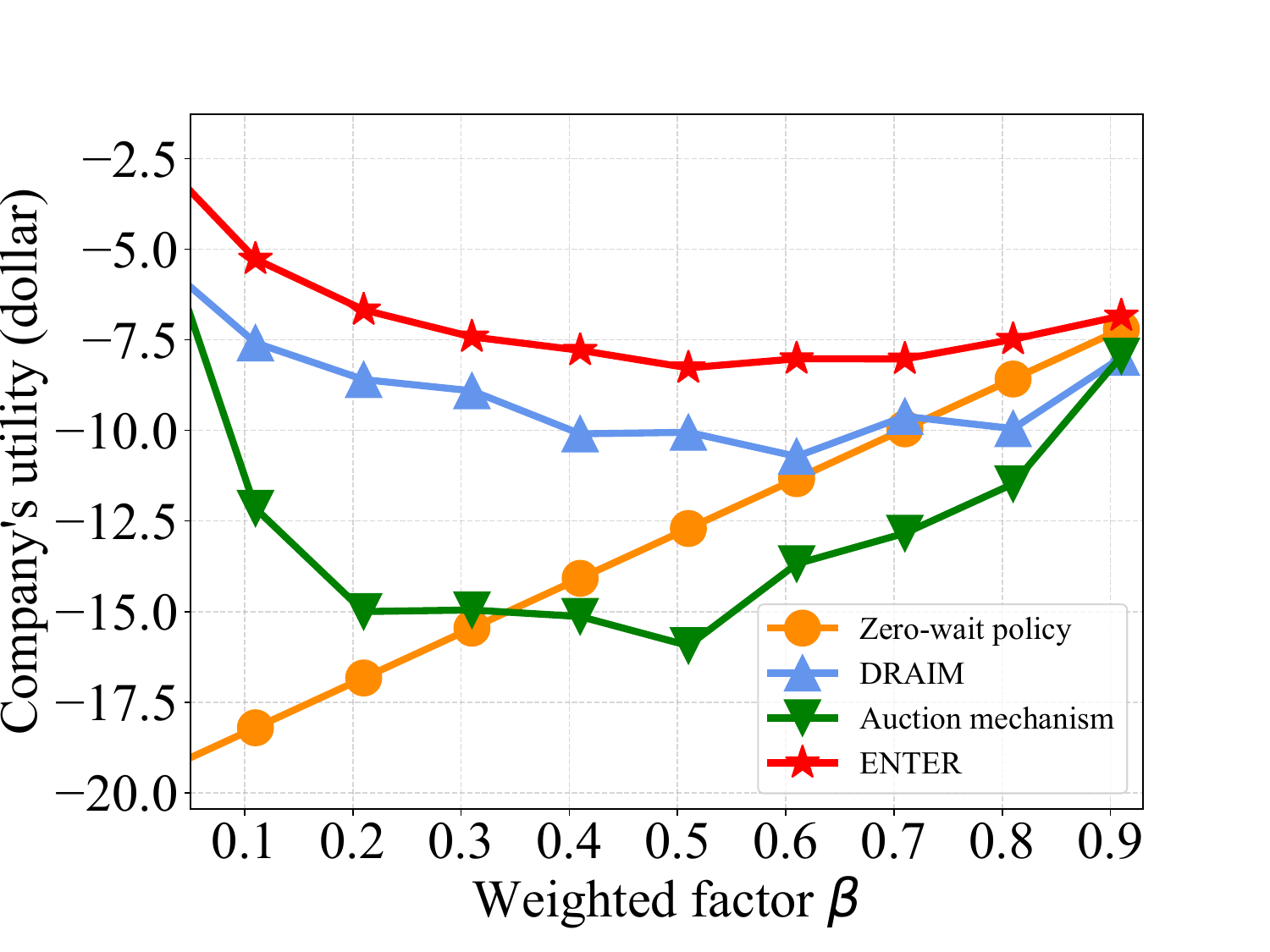}
    \caption{\rev{The {\cp}'s \payoff~\emph{vs.} the weighted factor $\Weightedfactor$ under different mechanisms.}}
    \vspace{-0.2cm}
    \label{fig: average cost comparison}
\end{figure}

We first conduct experiments to evaluate the impact of the weighted factor $\Weightedfactor$ on the {\cp}'s \payoff~as shown in Fig.~\ref{fig: average cost comparison}. In this figure, the x-axis represents the weighted factor $\Weightedfactor$, while the y-axis indicates the company's \payoff. For illustrative purposes, we set the number of vehicle types $\TypeNumber=4$ and randomly assign vehicle parameters using uniform distributions. 

Fig.~\ref{fig: average cost comparison} illustrates the \texttt{ENTER} mechanism's significant \payoff~increase, outperforming the zero-wait policy, DRAIM, and auction mechanism by \rev{$38.34\%$, $23.40\%$, and $43.91\%$}, respectively. The \payoff~of the company using the zero-wait policy increases linearly as it focuses solely on minimizing AoI loss, ignoring recruitment costs scaled by $1-\Weightedfactor$. Conversely, the other policies, which balance AoI loss with recruitment costs, produce arched cost profiles, \emph{i.e.,} first decrease and then increase. Notably, DRAIM and the auction mechanism's costs fluctuate due to a recruitment strategy based on an idealized \payoff~that assumes uniform vehicle parameters, overlooking actual variations that lead to cost inconsistencies.

\begin{figure}[h] 
  \centering
  \begin{minipage}{.66\linewidth} 
    \centering
    \includegraphics[width=\linewidth]{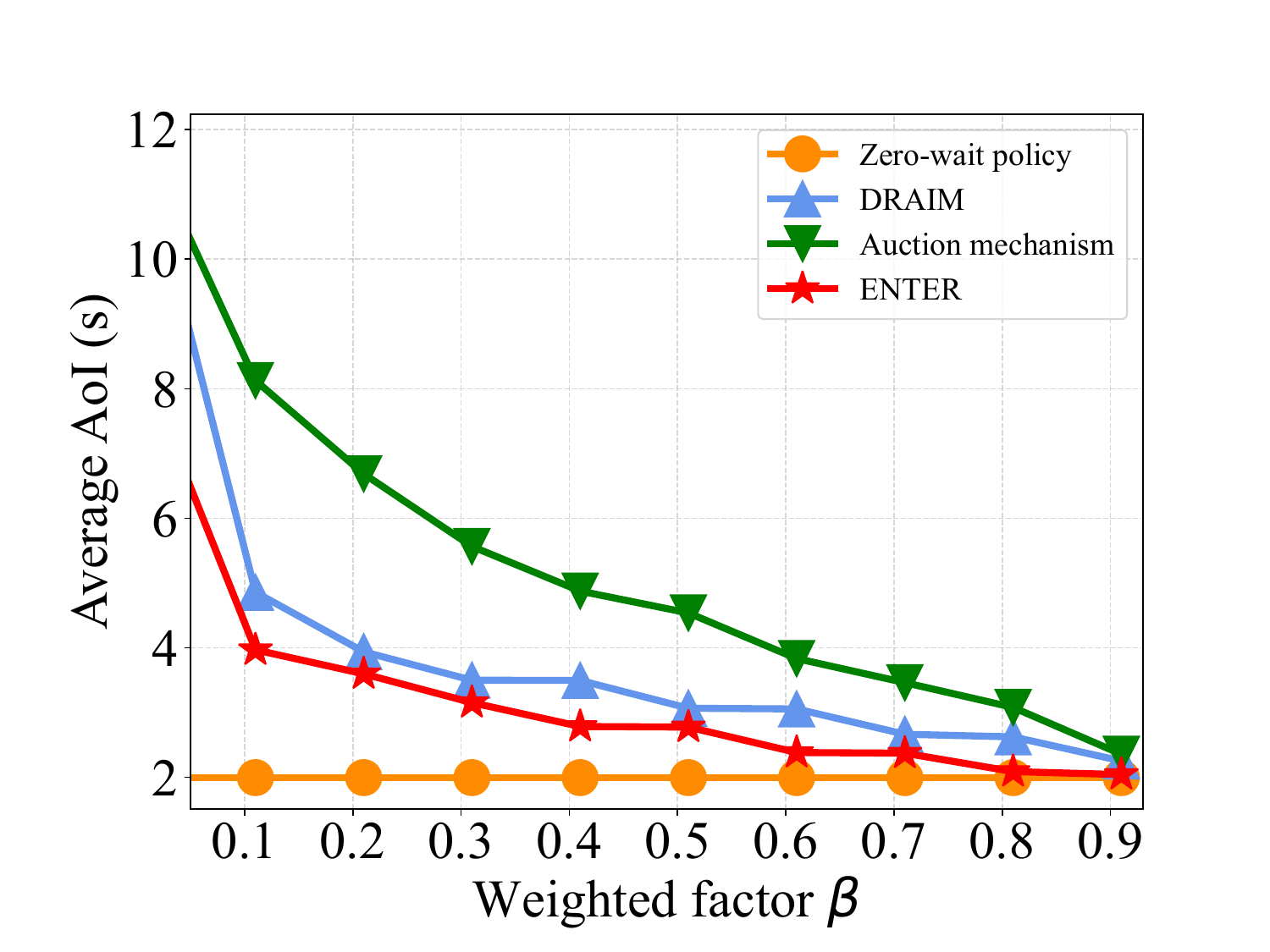} 
    \caption{\rev{The average AoI \emph{vs.} the weighted factor $\Weightedfactor$ under different mechanisms.}} 
    \label{subfig:weighted map freshness} 
  \end{minipage}
  \hfill 
  \begin{minipage}{.66\linewidth}
    \centering
    \includegraphics[width=\linewidth]{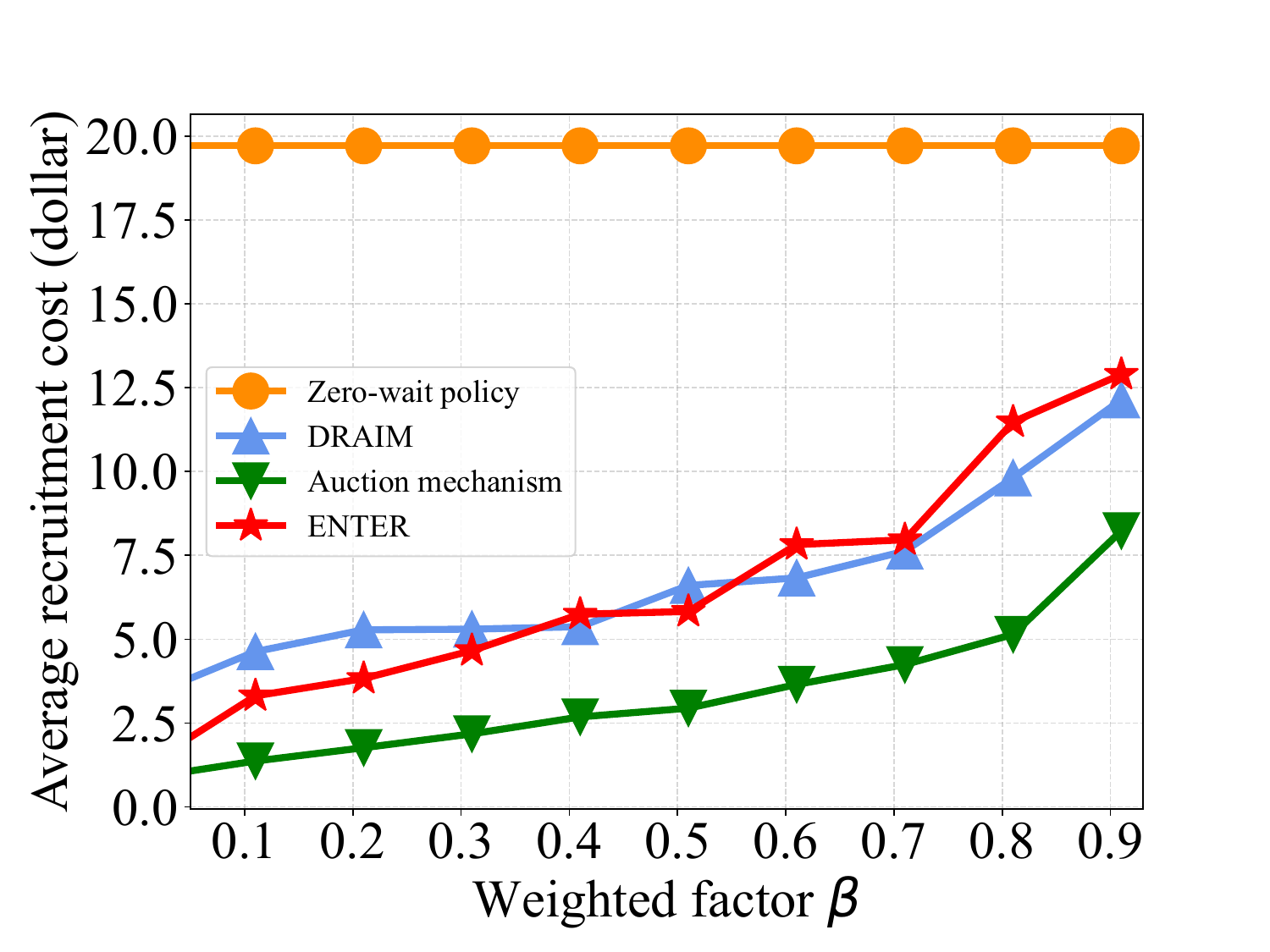} 
    \caption{\rev{The average recruitment cost \emph{vs.} the weighted factor $\Weightedfactor$ under different mechanisms.}} 
    \label{subfig:weighted cost} 
  \end{minipage}
\end{figure}

\rev{The experimental results of the weighted factor's impact on the average AoI and recruitment cost are illustrated in Fig.~\ref{subfig:weighted map freshness} and \ref{subfig:weighted cost}, respectively. 
The y-axes of Fig.~\ref{subfig:weighted map freshness} and \ref{subfig:weighted cost} indicate the average AoI and recruitment cost, respectively, while the x-axes represent the weighted factor. Fig.~\ref{subfig:weighted map freshness} and \ref{subfig:weighted cost} demonstrates notable
performance improvements and underscore the nuanced trade-offs that the \texttt{ENTER} mechanism offers in optimizing both AoI and recruitment costs within the context of our study.
First, it exhibits a $15.59\%$ reduction in the average AoI while keeping the recruitment cost at $94.91\%$ compared to DRAIM. Second, the \texttt{ENTER} mechanism saves $66.01\%$ of the average recruitment cost, albeit at the expense of a $48.37\%$ increase in the average AoI compared to the zero-wait policy. Last, the \texttt{ENTER} mechanism incurs a $106.34\%$ increase in the average recruitment cost, counterbalanced by a significant $37.68\%$ reduction in the average AoI compared to the auction mechanism. It is noteworthy that while the zero-wait policy and the auction mechanism may excel in specific isolated aspects, the \texttt{ENTER} mechanism ultimately optimizes the company's utility most effectively.}

\subsubsection{The impact of the number of vehicle types on the company's \payoff}
\begin{figure}[htbp]
    \centering
    \vspace{-0.4cm}
    \includegraphics[width=.66\linewidth]{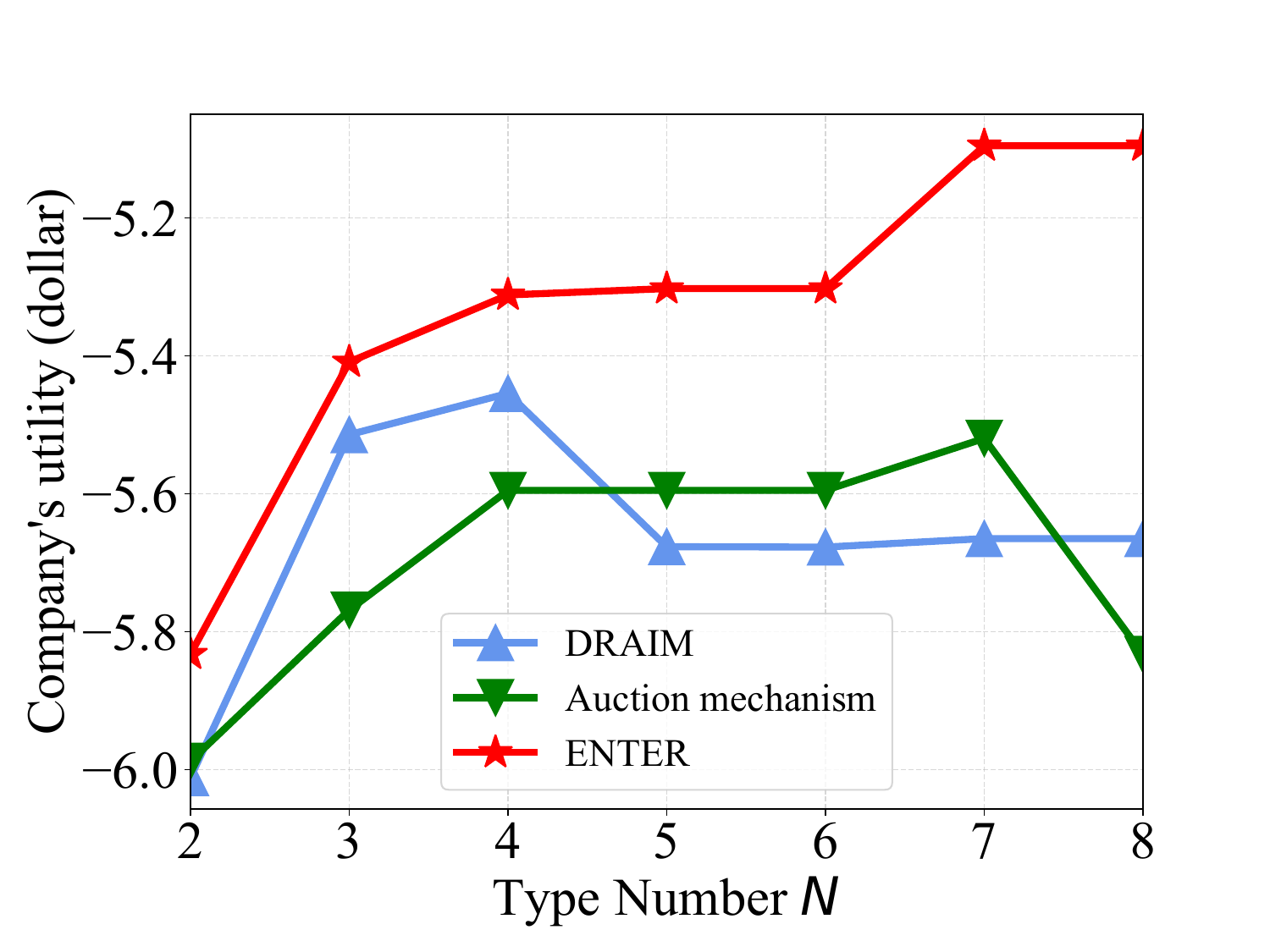}
    \caption{\rev{The {\cp}'s \payoff~\emph{vs.} the number of vehicle types under different mechanisms.}}
    \vspace{-0.2cm}
    \label{fig:average cost comparison on N}
\end{figure}

We further conduct numerical experiments to study the impact of the number of vehicle types on the {\cp}'s \payoff~under different mechanisms. Fig.~\ref{fig:average cost comparison on N} shows the result with the x-axis representing the number of vehicle types $\TypeNumber$ and the y-axis indicating the corresponding \payoff~incurred by the company. For illustrative purposes, we set the weighted factor $\Weightedfactor=0.1$. Note that in Fig.~\ref{fig:average cost comparison on N}, as $\TypeNumber$ increases by 1, we add one new randomly generated vehicle type to the existing fleet, with vehicle parameters \rev{$(\ArrivalprobN,\mathcal{C}_n,\mathcal{R}_n)$} drawn uniformly at random. This allows us to evaluate how the policies perform as vehicles' heterogeneity in HD map crowdsourcing increases in a stochastic manner. The weighted factor remains constant to isolate the impact of growing vehicle type heterogeneity.

Fig.~\ref{fig:average cost comparison on N} illustrates how the \texttt{ENTER} mechanism increases the company's \payoff~compared to baseline methods as the number of vehicle types grows. 
Our mechanism consistently outperforms the baselines for any number of types. Specifically, when there are eight types, our method improves the company's \payoff~by \rev{$5.83\%$ and $6.38\%$} versus the DRAIM and auction mechanisms, respectively. Notice that we omit the zero-wait policy, where the company's \payoff~is much less than others and monotonically decreases with the type number $\TypeNumber$.

The curves in Fig.~\ref{fig:average cost comparison on N} show sharp fluctuations due to the inclusion of vehicle types with relatively high or low parameters. For instance, the blue curve (DRAIM) shows a sharp drop when incorporating type 5 vehicles with low sensing \rev{capability mean} ($r_5=0.18$), compared to the mean value of $0.36$ for the first four vehicle types. However, DRAIM incorrectly assumes type 5 vehicles to be perfect ($\GSensingcap_5=1$). Conversely, the sudden rise in the red (the \texttt{ENTER} mechanism) and green (Auction mechanism) curves result from type 7 vehicles’ high sensing \rev{capability mean} ($r_7 = 0.45$) and arrival probability ($p_7 = 0.39$). Additionally, the infrequent arrivals of type 8 vehicles ($p_8=0.04$) mean the auction mechanism rarely recruits them, leading to significant AoI loss.

\begin{figure}[h] 
  \centering
  \begin{minipage}{.66\linewidth} 
    \centering
    \includegraphics[width=\linewidth]{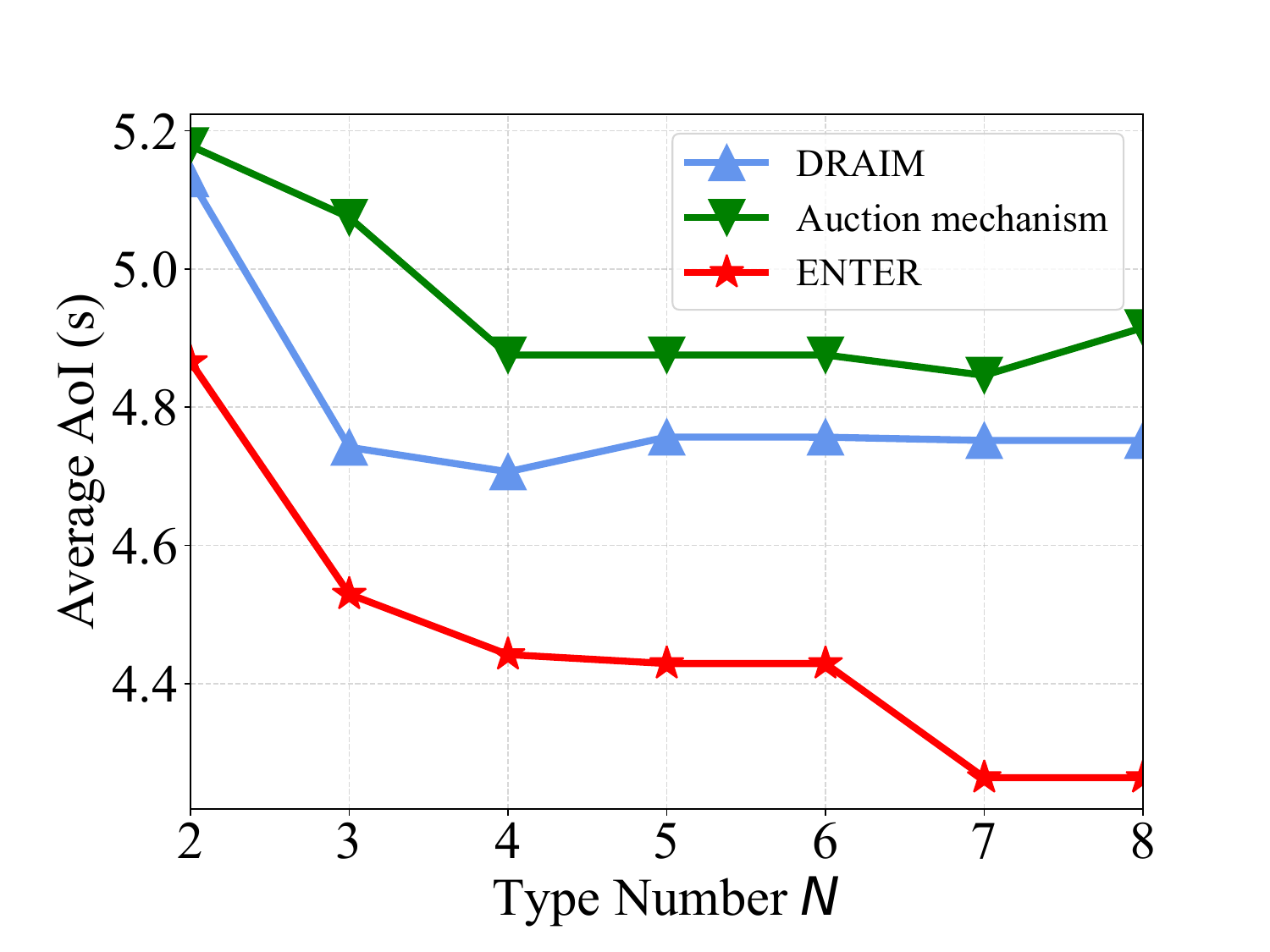} 
    \caption{\rev{The average AoI \emph{vs.} the number of vehicle types under different mechanisms.}} 
    \label{subfig:typeNum map freshness} 
  \end{minipage}
  \hfill 
  \begin{minipage}{.66\linewidth}
    \centering
    \includegraphics[width=\linewidth]{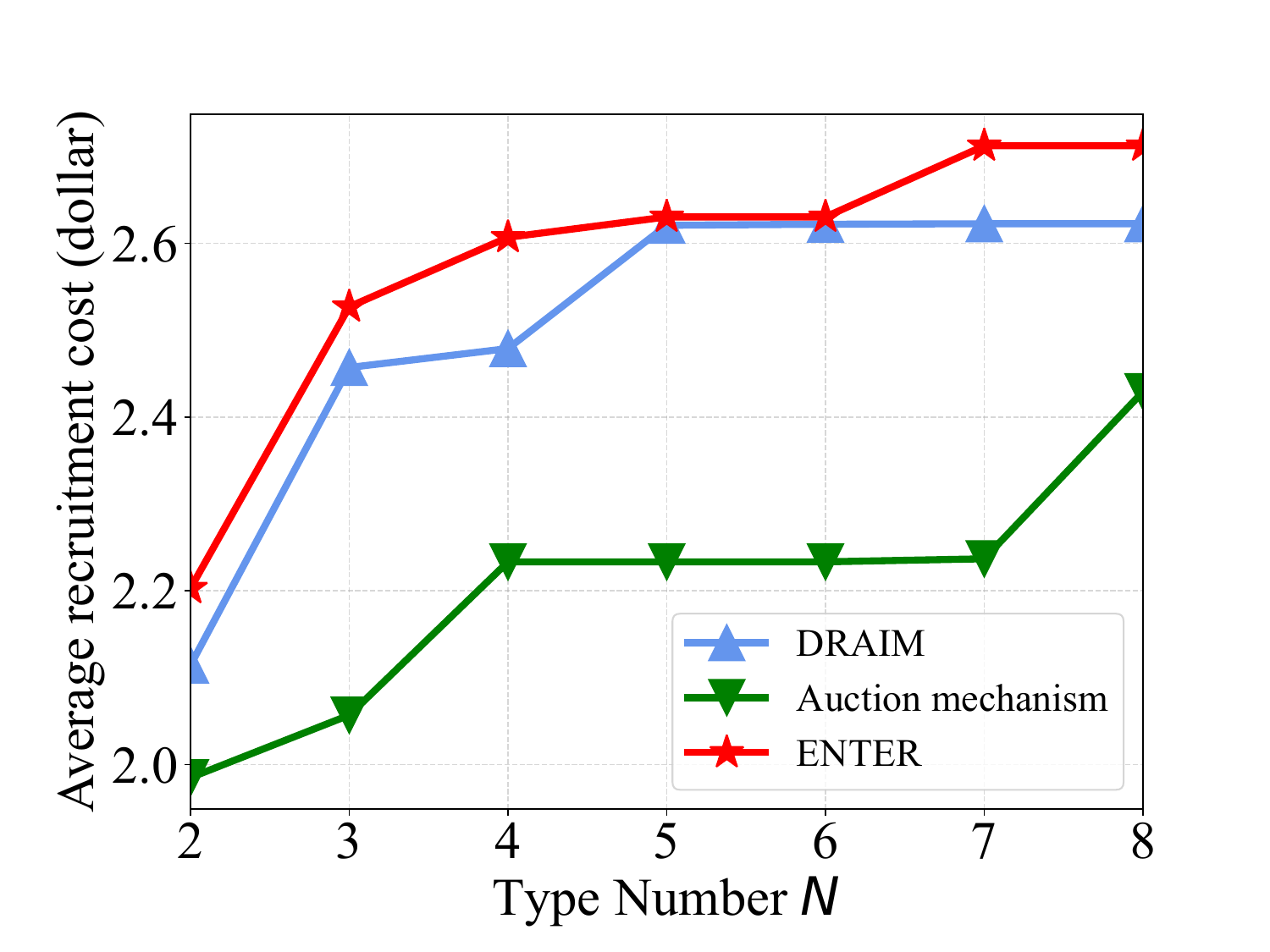} 
    \caption{\rev{The average recruitment cost \emph{vs.} the number of vehicle types under different mechanisms.}} 
    \label{subfig:typeNum cost} 
  \end{minipage}
\end{figure}

\rev{We examine the impact of the number of vehicle types on the average AoI and recruitment cost, as illustrated in Fig.~\ref{subfig:typeNum map freshness} and \ref{subfig:typeNum cost}. In these figures, the y-axes denote the average AoI and recruitment cost correspondingly, while the x-axes signify the varying number of vehicle types. It is important to note that we do not juxtapose these results with the average recruitment cost of the zero-wait policy, as it notably surpasses and consistently rises with an increase in vehicle types. 
As depicted in Fig.~\ref{subfig:typeNum cost}, the \texttt{ENTER} mechanism exhibits the highest average recruitment cost. However, except for achieving the highest utility, it attains the lowest average AoI compared to the DRAIM and auction mechanisms, as illustrated in Fig.~\ref{subfig:typeNum map freshness}. Notably, with eight vehicle types, the \texttt{ENTER} mechanism leads to a $3.42\%$ and $11.65\%$ increase in average recruitment cost compared to the DRAIM and auction mechanisms, respectively. However, it concurrently delivers reductions of $10.27\%$ and $13.24\%$ in the average AoI, respectively.}

In the next subsection, we will compare the bound-based RVI algorithm in the \texttt{ENTER} mechanism with the leading RVI-based algorithm to demonstrate its efficiency.

\subsection{Efficiency Comparison} \label{subsec: efficiency comparison}

In this subsection, we compare the proposed bound-based RVI algorithm in the \texttt{ENTER} mechanism to two baseline RVI-based algorithms in terms of computation time. \rev{It is worth noting that these algorithms solve the same MDP problem to ensure fairness.}

\begin{itemize}
    \item RVI algorithm: A classical dynamic programming method to solve the average cost MDP problem.
    \item Structural RVI (SRVI) algorithm \cite{hsu2019scheduling}: An improved RVI algorithm to reduce computational complexity based on the threshold-type structural property of the optimal policy, which is also state-of-the-art.
\end{itemize}

\begin{figure}[htbp]
    \centering
    \vspace{-0.4cm}
    \includegraphics[width=.66\linewidth]{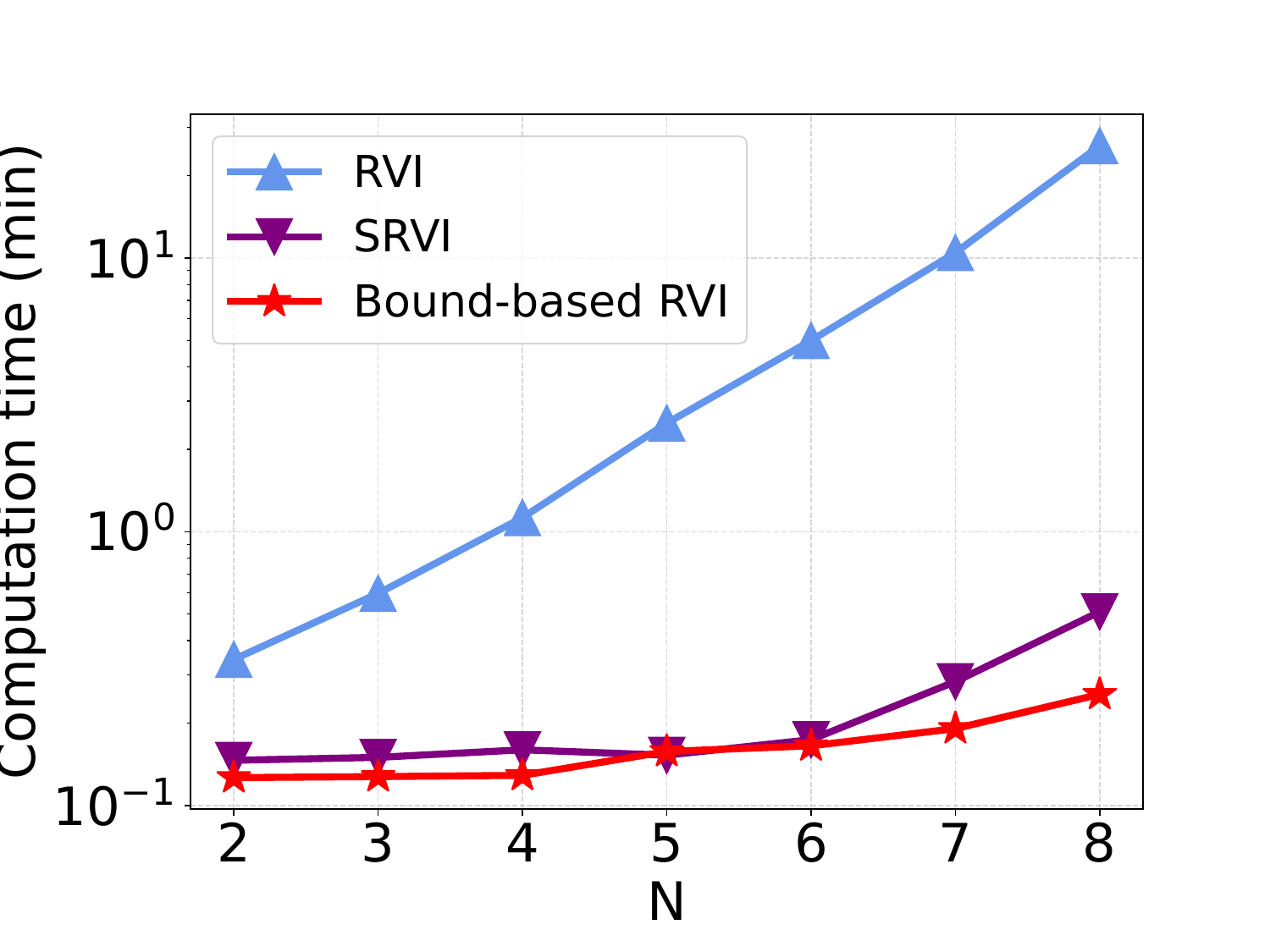}
    \caption{\rev{Comparison of computation time.} }
    \vspace{-0.2cm}
    \label{fig:efficiencycompare}
\end{figure}

Fig.~\ref{fig:efficiencycompare} compares the computation time for different numbers of vehicle types $\TypeNumber$. The x-axis corresponds to the number of vehicle types $\TypeNumber$, while the y-axis, logarithmically scaled, represents the computation time of different algorithms. This scaling is necessary due to the rapid increase in computation time with the exponentially growing feasible space in the RVI algorithm.  For illustration, we set the weighted factor  $\Weightedfactor=0.1$, the convergence tolerance $\epsilon=10^{-10}$, and the number of truncated states $\truncateNum=1000$. Additionally, we randomly set the vehicles’ parameters. To mitigate randomness and ensure reliability, we repeat the simulation five times and average the computation time across all repetitions.

Fig.~\ref{fig:efficiencycompare} shows that the bound-based RVI algorithm in the \texttt{ENTER} mechanism performs similarly to the SRVI algorithm when $\TypeNumber$ is small but significantly outperforms it as $\TypeNumber$ increases. On average, the bound-based RVI algorithm reduces computation time by \rev{$88.24\%$} compared to the RVI algorithm and by \rev{$18.91\%$} compared to the SRVI algorithm\footnote{Even for the 24 types of vehicles, the proposed Bound-based RVI algorithm converges faster more than three times than SRVI.}. Therefore, the \texttt{ENTER} mechanism, including the bound-based RVI algorithm, is better suited for dynamic environments with heterogeneous vehicles.

Fig.~\ref{fig:efficiencycompare} illustrates how the convergence time of each algorithm scales with the number of vehicle types. For RVI, the time increases sharply due to the expanding feasible set. The SRVI algorithm and our proposed algorithm experience a decrease and then an increase. This is because, for a small number of types, the iterative benefits of exploiting the threshold-type structure outweigh the growing feasible space, but this effect is eventually outweighed. However, our bound-based RVI algorithm alleviates the negative effect of its growth by reducing the feasible set. Overall, bound-based RVI maintains better scalability as the number of vehicle types increases.

In the next subsection, we will present a binary-type case for a deeper insight into the optimal policy, such as the impact of the vehicle's arrival probability on the optimal policy.

\subsection{Parameter Sensitivity}\label{subsec:para sensitivity}
In this subsection, we use a binary-type case to study the impact of vehicle parameters on the thresholds of the optimal policy. This analysis provides key insights that can inform more practical and complex scenarios.

Due to the complexity involved with multiple vehicle types and their parameters, as discussed in previous sections, analyzing the impact of these parameters on the optimal policy is challenging. Therefore, we simplify our analysis by focusing on the binary-type case. This approach not only makes the analysis more manageable but also effectively reveals the impact of these parameters and is practical in specific scenarios. Sensor configurations directly correlate with a vehicle's advanced driver-assistance systems (ADAS) and autonomous driving levels. Currently, most vehicles fall under Level 1 and 2 autonomy \cite{categorynow}. Consequently, we can consider only two types of vehicles, \Tl~and \Th, to gain deeper insight into the optimal policy. \rev{For clarity and simplification, we represent \Tl~and \Th~vehicles' arrival probabilities, constant operational costs, and constant sensing capabilities with $p_L$ and $p_H$, $\Lcost$ and $\Hcost$, as well as $\Lqual$ and $\Hqual$, respectively.} \Tl~vehicles have worse sensors than \Th~vehicles, statistically, implying that their {\sensingcaps} satisfy $\Lqual < \Hqual$.

Based on Theorem~\ref{thm:optimal policy}, we characterize four possible optimal policy structures of the binary-type case in Corollary~\ref{col:binary-type structure}. To illustrate, we define \None~as the action of no recruitment, \Ltype~as the action of only recruiting Low-type vehicles, \Htype~as the action of only recruiting High-type vehicles, and \Both~as the action of recruiting both types. 

\begin{corollary}\label{col:binary-type structure}
    There exists a threshold-type age-dependent stationary deterministic optimal policy $\policy^*$ of the MDP $\MDP$. Depending on the parameter setting, such an optimal policy falls into one of the following four structures:
    \begin{enumerate}
        \item \textbf{LH structure} (in Fig.~\ref{fig: LH structure}): if Condition~\eqref{cond:LH} is satisfied, the optimal policy includes actions in the order of \None, \Ltype, \Htype, \Both, with three thresholds satisfying $1 \leq \threshold_{N \to L} \leq \threshold_{L \to H} \leq \threshold_{H \to B}\in\Nsetp$.
        \begin{eqnarray} \label{cond:LH}
            \frac{1-\GSuccessprobN_H}{1-\GSuccessprobN_L}<\frac{\MarCosteffectivenessN_{O,L}}{\MarCosteffectivenessN_{O,H}}<1
        \end{eqnarray}
        \item \textbf{HL structure} (in Fig.~\ref{fig: HL structure}): if Condition~\eqref{cond:HL} is satisfied, the optimal policy includes actions in the order of \None, \Htype, \Ltype, \Both, with three thresholds satisfying $1\leq \threshold_{N\to H} \leq \threshold_{H\to L} \leq \threshold_{L\to B}\in \Nsetp$.
        \begin{eqnarray}\label{cond:HL}
            1<\frac{\MarCosteffectivenessN_{O,L}}{\MarCosteffectivenessN_{O,H}}<\frac{1-\GSuccessprobN_H}{1-\GSuccessprobN_L}
        \end{eqnarray}
        \item \textbf{None-L structure} (in Fig.~\ref{fig: None-L structure}): if Condition~\eqref{cond:None-L} is satisfied, the optimal policy includes actions in the order of \None, \Htype, \Both, with two thresholds satisfying $1 \leq \threshold_{N\to H} \leq \threshold_{H\to B} \in \Nsetp$.
        \begin{eqnarray}\label{cond:None-L}
            \frac{\MarCosteffectivenessN_{O,L}}{\MarCosteffectivenessN_{O,H}} \geq \max\left\{1,\frac{1-\GSuccessprobN_H}{1-\GSuccessprobN_L}\right\}
        \end{eqnarray}
        \item \textbf{None-H structure} (in Fig.~\ref{fig: None-H structure}): if Condition~\eqref{cond:None-H} is satisfied, the optimal policy includes actions in the order of \None, \Ltype, \Both, with two thresholds satisfying $1 \leq \threshold_{N\to L} \leq \threshold_{L\to B} \in \Nsetp$.
        \begin{eqnarray}\label{cond:None-H}
            \frac{\MarCosteffectivenessN_{O,L}}{\MarCosteffectivenessN_{O,H}}<\min\left\{1,\frac{1-\GSuccessprobN_H}{1-\GSuccessprobN_L}\right\}
        \end{eqnarray}
    \end{enumerate}
    
    \begin{figure}
     \centering
     \begin{subfigure}[b]{0.24\textwidth}
         \centering
         \includegraphics[width=\textwidth]{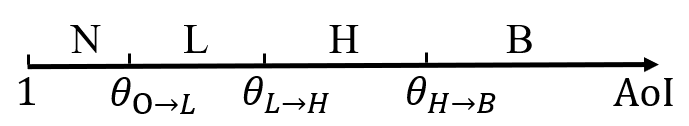}
         \caption{LH structure}
         \label{fig: LH structure}
     \end{subfigure}
     \hfill
     \begin{subfigure}[b]{0.24\textwidth}
         \centering
         \includegraphics[width=\textwidth]{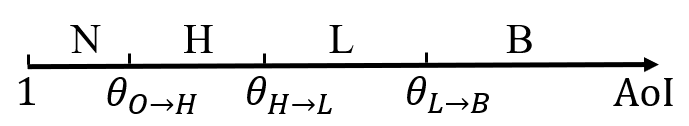}
         \caption{HL structure}
         \label{fig: HL structure}
     \end{subfigure}
     \hfill
     \begin{subfigure}[b]{0.24\textwidth}
         \centering
         \includegraphics[width=\textwidth]{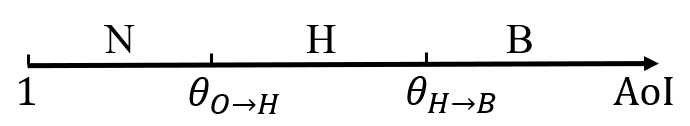}
         \caption{None-L structure}
         \label{fig: None-L structure}
     \end{subfigure}
     \begin{subfigure}[b]{0.24\textwidth}
         \centering
         \includegraphics[width=\textwidth]{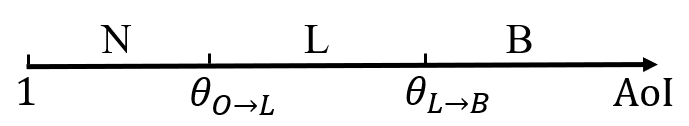}
         \caption{None-H structure}
         \label{fig: None-H structure}
     \end{subfigure}
        \caption{Four possible structures of the optimal policy.}
        \label{fig: possible structures}
        \vspace{-0.4cm}
\end{figure}
\end{corollary}

Corollary~\ref{col:binary-type structure} outlines four structures tailored to various real-world scenarios. The proof of Corollary~\ref{col:binary-type structure} is detailed in Appendix I. 
The LH and HL structures apply when \tl~and \th~vehicles have comparable cost-effectiveness. That means the ratio of the marginal cost-effectiveness of recruiting \th~ only  (\emph{i.e.,} action \Htype) versus recruiting \tl~ only (\emph{i.e.,} action \Ltype) vehicles from not recruiting any vehicles (\emph{i.e.,} action \None) falls between $(1-\GSuccessprobN_{H})/(1-\GSuccessprobN_{L})$ and $1$. Specifically, the LH structure is adopted for the slight cost-effectiveness superiority of \tl~vehicles, while the HL structure is used when \th~vehicles are more cost-effective. The None-L structure is chosen when recruiting \th~vehicles far outweighs recruiting \tl~vehicles, as the ratio $\MarCosteffectivenessN_{O,L}/\MarCosteffectivenessN_{O,H}$ surpasses both $(1-\GSuccessprobN_{H})/(1-\GSuccessprobN_{L})$ and $1$. Conversely, the None-H structure indicates that \tl~vehicles are far more cost-effective. 

\begin{figure}[h]
    \centering
    \vspace{-0.4cm}
    \includegraphics[width = .75\linewidth]{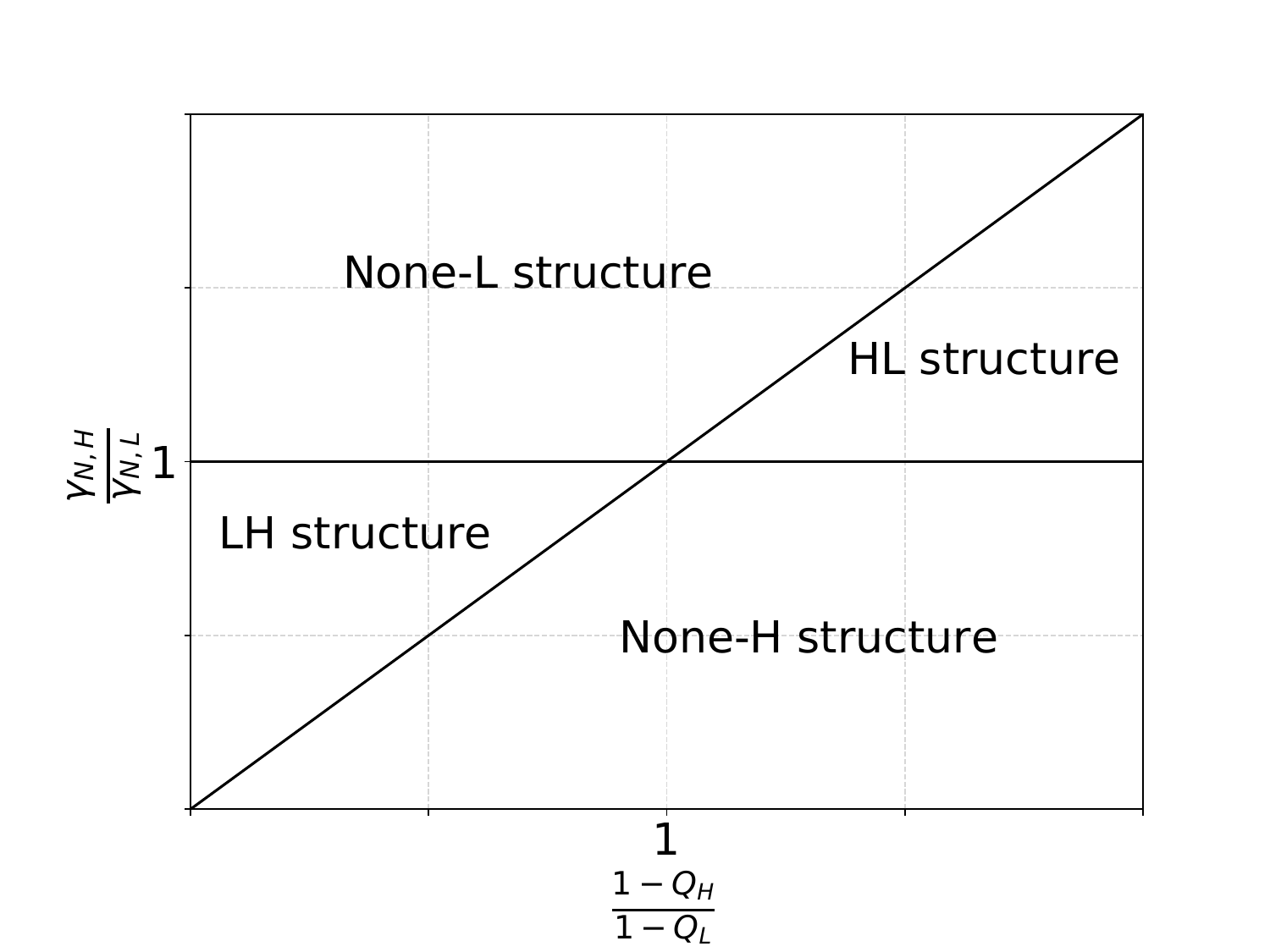}
    \caption{The distribution of the optimal policy structures.} \label{fig: dis of optimal}
    \vspace{-0.2cm}
\end{figure}

To illustrate the relationship between the optimal policy structure, success probability, and marginal cost-effectiveness, we plot the distribution of the optimal policy structures in Fig.~\ref{fig: dis of optimal}. Here, the x-axis represents the ratio of the reciprocal success probabilities between actions \Ltype~and \Htype, while the y-axis shows the marginal cost-effectiveness ratio of recruiting \th~over \tl~vehicles without any prior recruitment.

Next, we investigate the impact of vehicle parameters, such as the {\arrivalprob}, on the thresholds of the optimal policy. By uncovering these connections, we reveal the underlying rationale for the optimal policy, which can apply to more practical and complex scenarios. 

\subsubsection{The impact of {\arrivalprobs}} \label{subsec:impact of arrival prob}
\begin{figure}
\vspace{-0.4cm}
     \centering
     \begin{subfigure}[b]{0.4\textwidth}
         \centering
         \includegraphics[width=.9\linewidth]{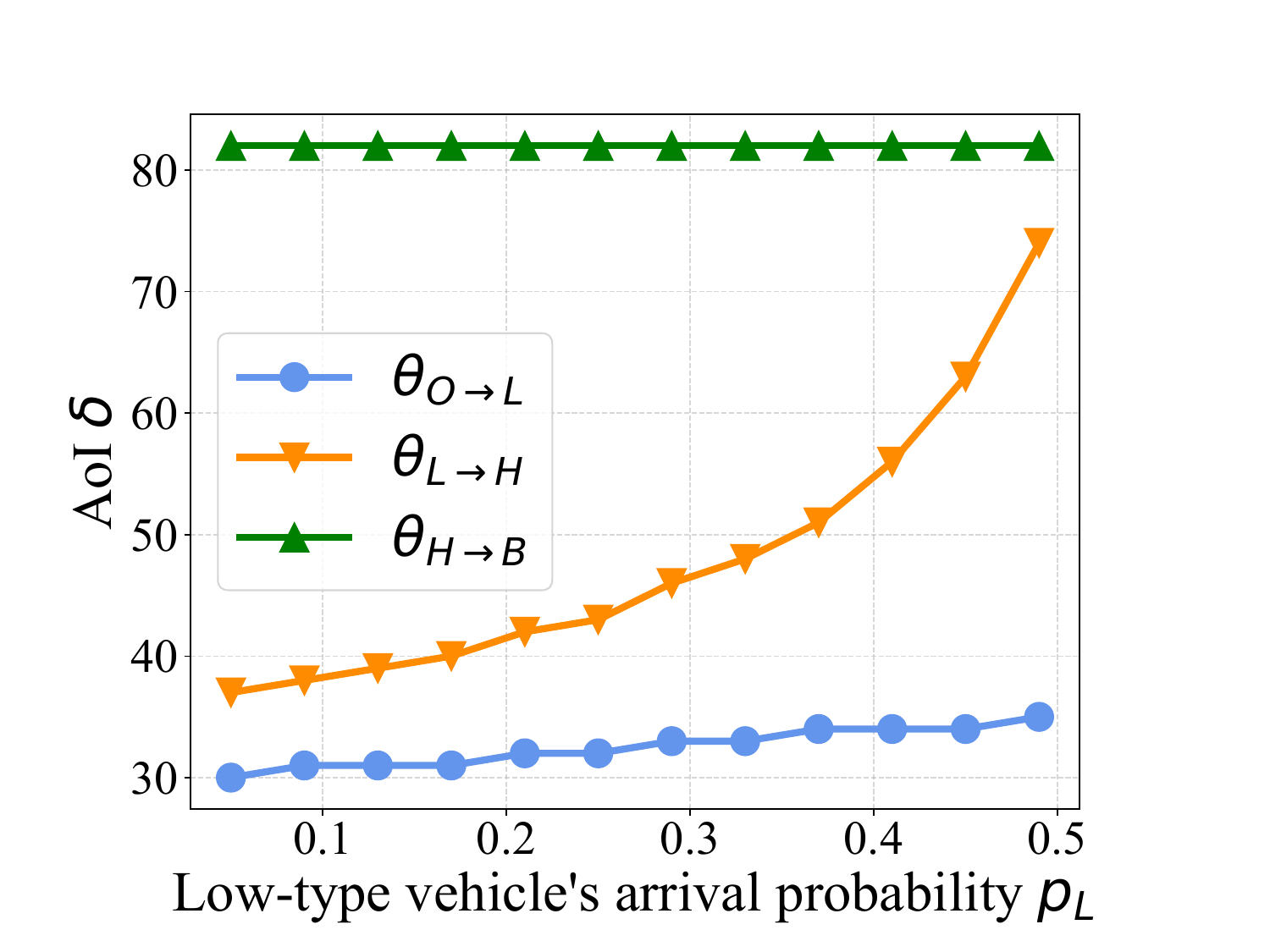}
         \caption{}
         \label{fig: p_L}
     \end{subfigure}
     \hfill
     \begin{subfigure}[b]{0.4\textwidth}
         \centering
         \includegraphics[width=.9\linewidth]{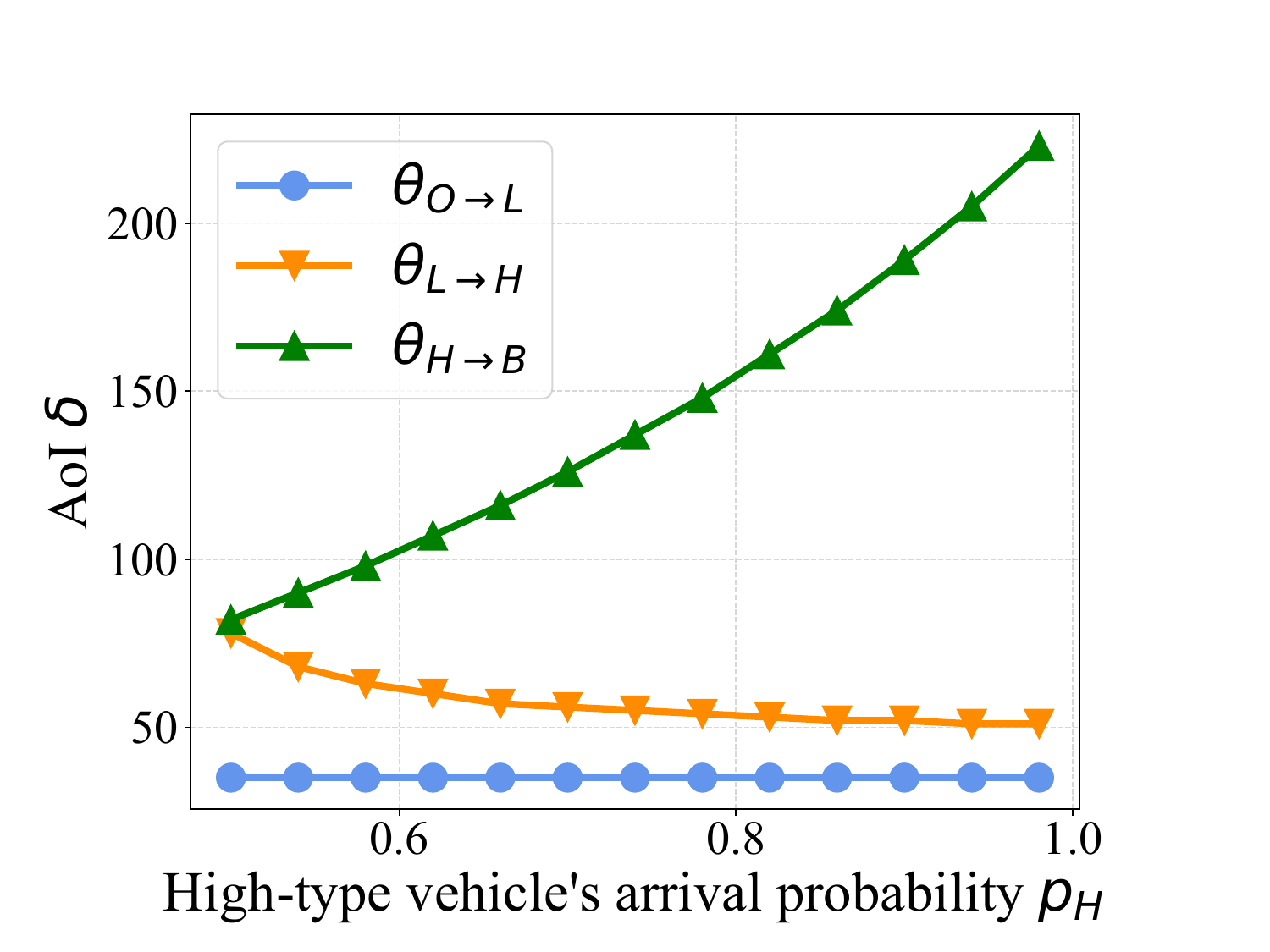}
         \caption{}
         \label{fig: p_H}
     \end{subfigure}
    \caption{Threshold vs. arrival probability in LH structure.}
    \label{fig: arrival prob}
    \vspace{-0.4cm}
\end{figure}

To examine the impact of {\arrivalprobs} on the optimal policy, we conducted experiments with vehicle parameters where \tl~and \th~vehicles have comparable cost-effectiveness (see Table~\ref{tab:vp LH structure}). This led to the adoption of the LH structure as the optimal policy. Fig.~\ref{fig: arrival prob} shows the numerical results. Here, Figs.~\ref{fig: p_L} and \ref{fig: p_H} plot arrival probabilities of \Tl~and \Th~vehicles on the x-axis against AoI on the y-axis, respectively. Besides, we set the weighted factor $\Weightedfactor=0.0001$ in the current and following experiments. Note that the blue curve ($\threshold_{O\to L}$) in Fig.~\ref{fig: p_L} appears uneven because the threshold must align with the AoI, which is an integer.

\begin{table}[h]
    \centering
    \begin{tabular}{ccccccc}
        \hline
        Experiment & $p_L$ & $p_H$ & $c_L$ & $c_H$ & $r_L$ & $r_H$ \\
        \hline
        Fig.~\ref{fig: p_L} & (0$\sim$0.5) & 0.5 & 2 & 2.5 & 0.6 & 0.7\\
        Fig.~\ref{fig: p_H} & 0.5 & (0.5$\sim$1) & 2 & 2.5 & 0.6 & 0.7\\
        \hline
    \end{tabular}
    \caption{Vehicle parameter setting for LH structure.}
    \vspace{-0.2cm}
    \label{tab:vp LH structure}
\end{table}

Fig.~\ref{fig: p_L} demonstrates that the threshold $\threshold_{N\to L}$ rises with the increasing arrival probability $\GArrivalprob_L$ of \Tl~vehicles (blue curve). This phenomenon supports the intuition that more frequent arrivals prompt recruitment under larger AoI to reduce costs. Conversely, Fig.~\ref{fig: p_H} shows that the threshold $\threshold_{L\to H}$ falls as the arrival probability $\GArrivalprob_H$ of \Th~vehicles increases (orange curve) due to the declined marginal cost-effectiveness $\MarCosteffectivenessN_{L, H}$. This decrease makes it more cost-effective to switch from action \Ltype~to action \Htype. Furthermore, these changes in marginal cost-effectiveness account for variations in the other curves in Fig.~\ref{fig: arrival prob}. Specifically, the increased marginal cost-effectiveness raises the corresponding threshold, while the decreased cost-effectiveness lowers it. Based on this analysis, we have the following observation.

\begin{obs}
    When vehicles arrive more frequently, the company recruits them at a smaller AoI if the marginal cost-effectiveness of recruiting these vehicles decreases.
\end{obs}

Notice that the rise in $\threshold_{N\to L}$ in Fig.~\ref{fig: p_L} is attributed to a higher success probability, offering a smaller expected AoI as vehicle arrivals become more frequent, despite stable marginal cost-effectiveness.

Due to space constraints, we have included the analysis of the impact of the vehicle’s {\sensingcap} and {\operationalcost} on the policy thresholds in Appendix J.

\section{Conclusion and Future work}

This paper presents the \texttt{ENTER} mechanism, which maximizes the company's \payoff~with vehicles of varying arrival probabilities, sensing capabilities, and operational costs. Considering the dynamics of AoI, we solve the company's \payoff~maximization problem by formulating an MDP with infinite states.  To address the challenge of solving an MDP with an infinite-state space, we introduced a truncated MDP. A key part of this approach involved developing a criterion to determine the optimal truncation number. The corresponding truncated MDP is solvable using RVI-based algorithms and crucially retains the same thresholds in the optimal policy as the original.
Building on this foundation, we developed a threshold-type age-dependent optimal policy for the MDP $\MDP$. To efficiently compute this policy, we introduce the bound-based RVI algorithm for efficient computation through the optimal policy structure and derived threshold upper bounds. Experimentally, the \texttt{ENTER} mechanism achieves a significant $44.92\%$ improvement in the company's \payoff~and cuts the computation time by $30.22\%$ compared to state-of-the-art algorithms.

Several promising avenues exist for future research. First, transitioning our model to a continuous-time framework with broader distributions would enhance real-world applicability. Second, investigating scenarios with multiple competing companies using a duopoly model would add complexity and depth. Third, integrating vehicle parameter predictions would lend additional realism. Overall, this work presents a foundation for optimally managing crowdsourced fleets, and opportunities remain to expand the scope and applicability of the approach.

\bibliographystyle{IEEEtran}
\bibliography{IEEEabrv,ref}

\begin{IEEEbiography}[{\includegraphics[width=1in,height=1.25in,clip,keepaspectratio]{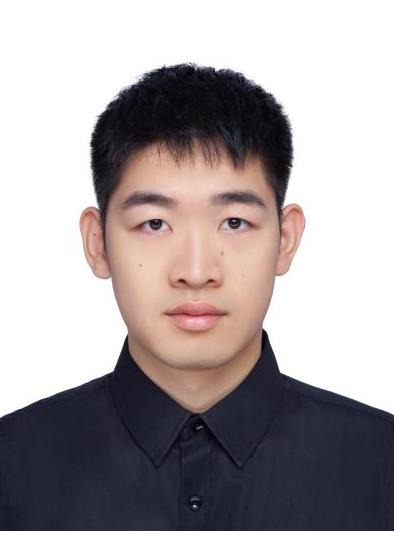}}]{Wentao Ye}
 received the BEng degree in Computer Science from the University of Electronic Science and Technology of China in 2021. He is currently working toward PhD degree with the Chinese University of Hong Kong, Shenzhen. Wentao Ye is with the Shenzhen Institute of Artificial Intelligence and Robotics for Society, School of Science and Engineering, The Chinese University of Hong Kong, Shenzhen, Guangdong, 518172, P.R. China. His research interests include crowdsourcing and mechanism design, with a current interest in mechanism design in LLM crowdsourcing systems.
\end{IEEEbiography}

\begin{IEEEbiography}[{\includegraphics[width=1in,height=1.25in,clip,keepaspectratio]{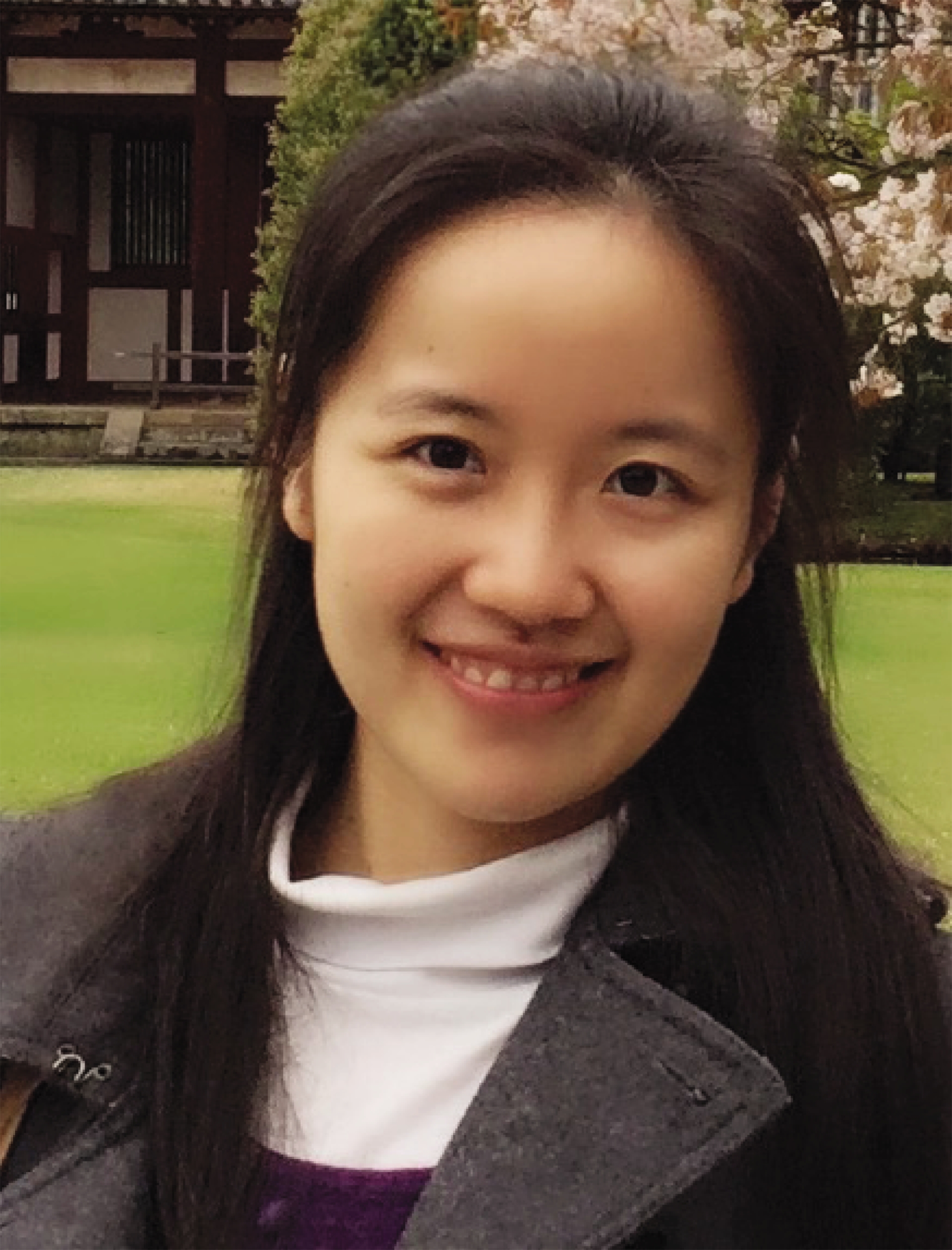}}]{Yuan Luo}
(S’10–M’16–SM’24) received the Ph.D. degree from The Chinese University of Hong Kong in 2015. She was a Visiting Scholar with University of California, Berkeley, USA, from 2014 to 2015, a Postdoctoral Researcher with The Chinese University of Hong Kong from 2015 to 2017, and a Research Fellow with Imperial College London, from 2017 to 2021. She is currently an Assistant Professor with the School of Science and Engineering, The Chinese University of Hong Kong, Shenzhen, China. Her research interests include Artificial Intelligence, Mechanism Design and Low-Latency Transmission. She is a recipient of the Best Paper Award in the IEEE WiOpt 2014.
\end{IEEEbiography}

\begin{IEEEbiography}[{\includegraphics[width=1in,height=1.25in,clip,keepaspectratio]{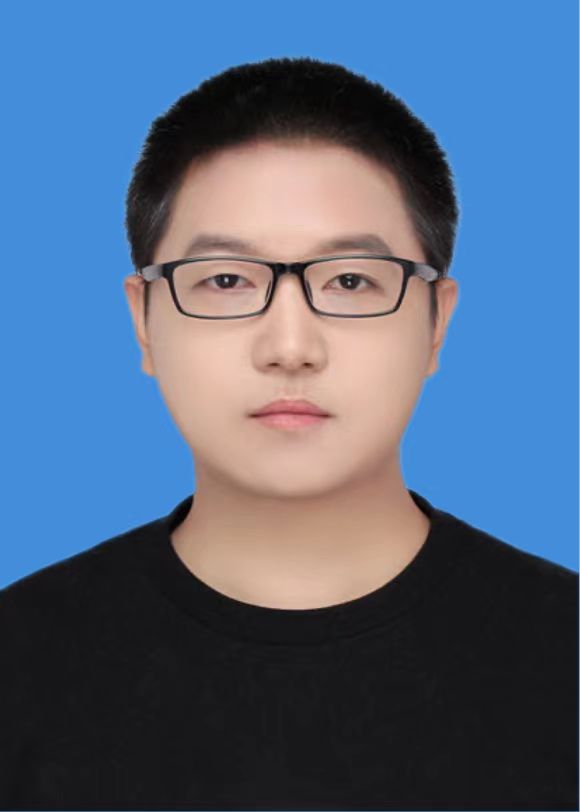}}]
{Bo Liu}(M'23) received the B.Eng. degree and M.Eng. degree from Wuhan University of Technology, Wuhan, China, in 2014 and 2017, respectively, and the Ph.D. degree in Electrical and Electronic Engineering at the University of Manchester, Manchester, U.K, in 2021. Currently, he is an associate research fellow in Shenzhen Institute of Artificial Intelligence and Robotics for Society, and his research interests include distributed optimization, distributed machine learning, and their applications.\end{IEEEbiography}

\begin{IEEEbiography}[{\includegraphics[width=1in,height=1.25in,clip,keepaspectratio]{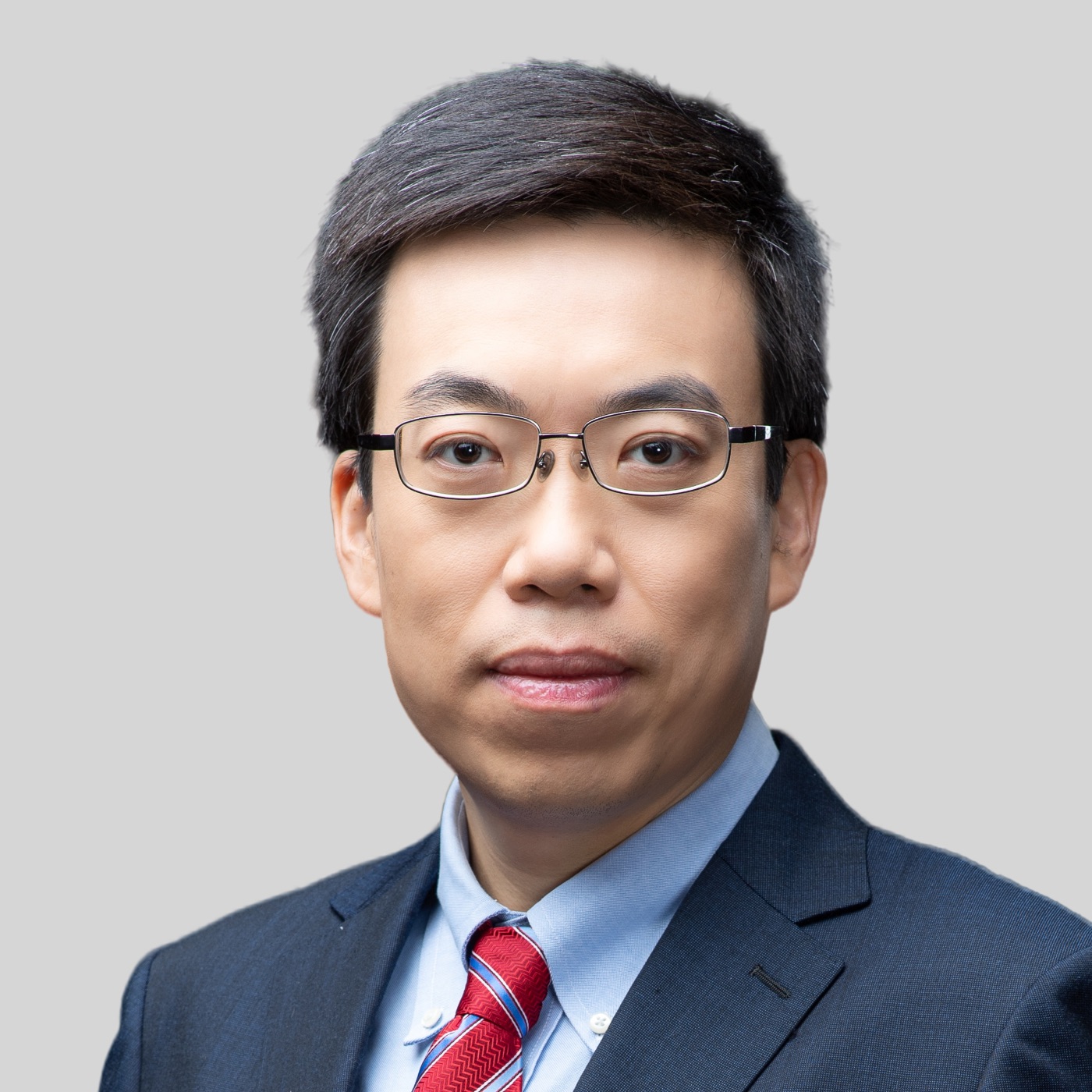}}]{Jianwei Huang} is a Presidential Chair Professor and Associate Vice President (Institutional Development) of the Chinese University of Hong Kong, Shenzhen, and the Associate Director of Shenzhen Institute of Artificial Intelligence and Robotics for Society. He received his Ph.D. from Northwestern University in 2005 and worked as a Postdoc Research Associate at Princeton University during 2005-2007. His research interests are network optimization and economics, with applications in communication networks, energy networks, data markets, and crowd intelligence. He has published 350+ papers in leading international venues, with a Google Scholar citation of 17,000+ and an H-index of 68. He has co-authored 11 Best Paper Awards, including the 2011 IEEE Marconi Prize Paper Award in Wireless Communications. He has co-authored seven books, including the textbook "Wireless Network Pricing." He has been an IEEE Fellow, an IEEE ComSoc Distinguished Lecturer, a Clarivate Web of Science Highly Cited Researcher, and an Elsevier Most Cited Chinese Researcher. He is the Editor-in-Chief of IEEE Transactions on Network Science and Engineering and Associate Editor-in-Chief of the IEEE Open Journal of the Communications Society.
\end{IEEEbiography}

\vspace{11pt}

\newpage
\newpage

\end{document}